\documentclass[]{spie}  

\usepackage{amsmath,amsfonts,amssymb}
\usepackage{graphicx}
\usepackage{algorithm2e}
\usepackage[colorlinks=true, allcolors=blue]{hyperref}
\newcommand{\arcsecond}{$^{\prime\prime}$}

\title{SCALES for Keck: Optical Design}

\author[a]{Renate Kupke}
\author[a]{R. Deno Stelter}
\author[b]{Amirul Hasan}
\author[c]{Arun Surya}
\author[a]{Isabel Kain}
\author[a]{Zackery Briesemeister}
\author[a]{Jialin Li}
\author[a]{Phil Hinz}
\author[a]{Andrew Skemer}
\author[a]{Benjamin Gerard}
\author[a]{Daren Dillon}
\author[a]{Christopher Ratliff}
\affil[a]{University of California Santa Cruz, 1156 High St, Santa Cruz, USA}
\affil[b]{Indian Institute for Astrophysics, Bengaluru, India}
\affil[c]{Tata Institute of Fundamental Research, Mumbai, India}

\authorinfo{Further author information: (Send correspondence to R. Kupke)\\R. Kupke: E-mail: rkupke@ucsc.edu }

\begin{document} 
\maketitle

\begin{abstract}
SCALES is a high-contrast, infrared coronagraphic imager and integral field spectrograph (IFS) to be deployed behind the W.M. Keck Observatory adaptive optics system. A reflective optical design allows diffraction-limited imaging over a large wavelength range (1.0 - 5.0 $\mu$m). A microlens array-based IFS coupled with a lenslet reformatter ("slenslit") allow spectroscopy at both low (R = 35 - 250) and moderate (R = 2000 - 6500) spectral resolutions. The large wavelength range, diffraction-limited performance, high contrast coronagraphy and cryogenic operation present a unique optical design challenge. We present the full SCALES optical design, including performance modeling and analysis and manufacturing. 
\end{abstract}


\section{INTRODUCTION}
\label{sec:intro}  

SCALES (Slicer Combined with Array of Lenslets for Exoplanet Spectroscopy) is a mid-infrared coronagraphic imager and integral field spectrograph (IFS) for the Keck telescope. Coupled with Keck Adaptive Optics (AO), it will enable a number of exciting science cases, including spectroscopic characterization of self-luminous exoplanets which are colder and lower mass than are currently accessible to existing high contrast instruments.  A full description of the SCALES' science case and update on instrument development can be found in several references\cite{DenoSCALES2020,Andy2022}.

The optical design of SCALES must preserve the low wavefront error and excellent image quality provided by the Keck AO system while of course maximizing throughput. The imager will operate from 1.0 - 5.3 $\mu$m with a selection of up to sixteen photometric and narrow bandpass filters, while the coronagraphically-fed integral field spectrograph will disperse in both low and medium spectral resolutions over a 2.0 - 5.3 $\mu$m bandpass. Both the imager and IFS will utilize 18 $\mu$m pixel H2RG (2048x2048) detectors.

\begin{figure}
    \centering
    \includegraphics[width=0.9\textwidth]{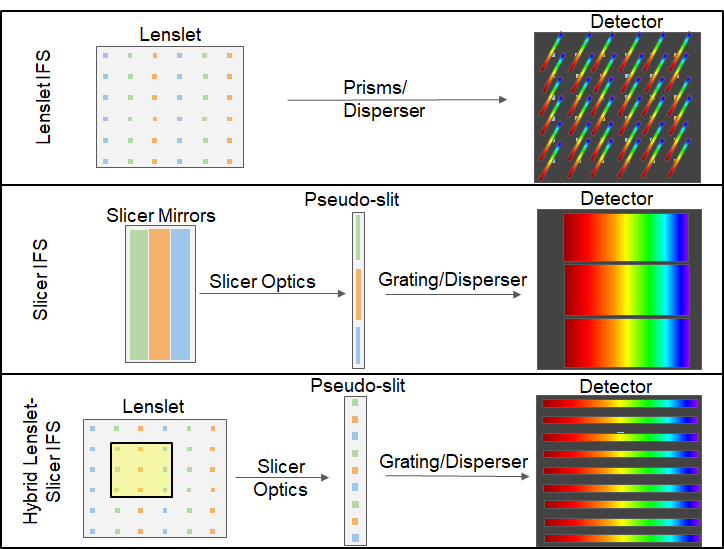}
    \caption{Cartoon of lenslet array IFS (top), slicer IFS (middle) and "Slenslit" (bottom) IFS spectral formats.}
    \label{fig:IFStypes}
\end{figure}

As with other adaptive optics-enabled integral field spectrographs, the SCALES design utilizes a microlens array integral field unit (IFU). This architecture has the advantage of spatially sampling the field of view before the spectrograph optics, which are often designed to be fast and have large field angles to accommodate spectral resolution and bandpass at the expense of wavefront quality. Each lenslet serves as a spatial pixel ("spaxel") which forms a micro-pupil image. This grid of pupil images is input for the spectrograph, and the spectra are interleaved by rotating the disperser prisms with respect to the lenslet array. This low spectral resolution configuration is represented by the cartoon shown in the top panel of Figure~\ref{fig:IFStypes}. 

SCALES has adopted a novel approach to obtaining moderate spectral resolutions over large bandpasses. We are implementing an image slicer developed by Stelter\cite{DenoSlenslit2021,DenoSlenslit2022} to reformat the pupil grid produced by the microlens array into a set of three interleaved pseudoslits, which are dispersed across the full length of the detector by gratings. The medium resolution configuration is shown on the lower panel of Figure~\ref{fig:IFStypes}.  We refer to the slicer as a "slenslit", as it combines the properties of an imager slicer IFU, a lenslet-based IFU, and a slit spectrograph. 
Requirements for the modes of operation, including field of view, spatial sampling, bandpass and spectral resolution are shown in Table {\ref{tab:SCALESReqSum}}. This paper starts with an overview of the optical design of the foreoptics, imager, and spectrograph. The slenslit\cite{DenoSlenslit2022} and imager\cite{Banyal2022} are described in detail in separate publications in this proceedings. We anticipate beginning fabrication of SCALES optical components in autumn 2022 with final alignment, testing and verification beginning in early 2024. 

\begin{table}[ht]
\caption{Summary of SCALES top-level requirements for each observing mode} 
\begin{center}       

\begin{tabular}{|r|l|l|l|}
\hline
 & \textbf{Imager} & \textbf{Low-Resolution} & \textbf{Medium-Resolution} \\
\hline
    Number of Spaxels & 2048x2048 & 108x108 & 17x18 \\
    Plate Scale & 0.06$^{\prime\prime}$ & 0.02$^{\prime\prime}$ & 0.02$^{\prime\prime}$ \\
    Field of View & 12.3$^{\prime\prime}$ $\times$ 12.3$^{\prime\prime}$ & 2.16$^{\prime\prime}$ $\times$ 2.16$^{\prime\prime}$ & 0.34$^{\prime\prime}$ $\times$ 0.36$^{\prime\prime}$ \\
    Bandpass & 1.0 - 5.3 $\mu$m & 2.0 - 5.3 $\mu$m & 2.0 - 5.3 $\mu$m \\
    Resolution & - & 35 - 250 & 2500 - 6500 \\
\hline
\end{tabular}

\end{center}
\label{tab:SCALESReqSum}
\end{table}

\section{OPTICAL DESIGN OVERVIEW}
\begin{figure}
    \centering
    \includegraphics[width=0.9\textwidth]{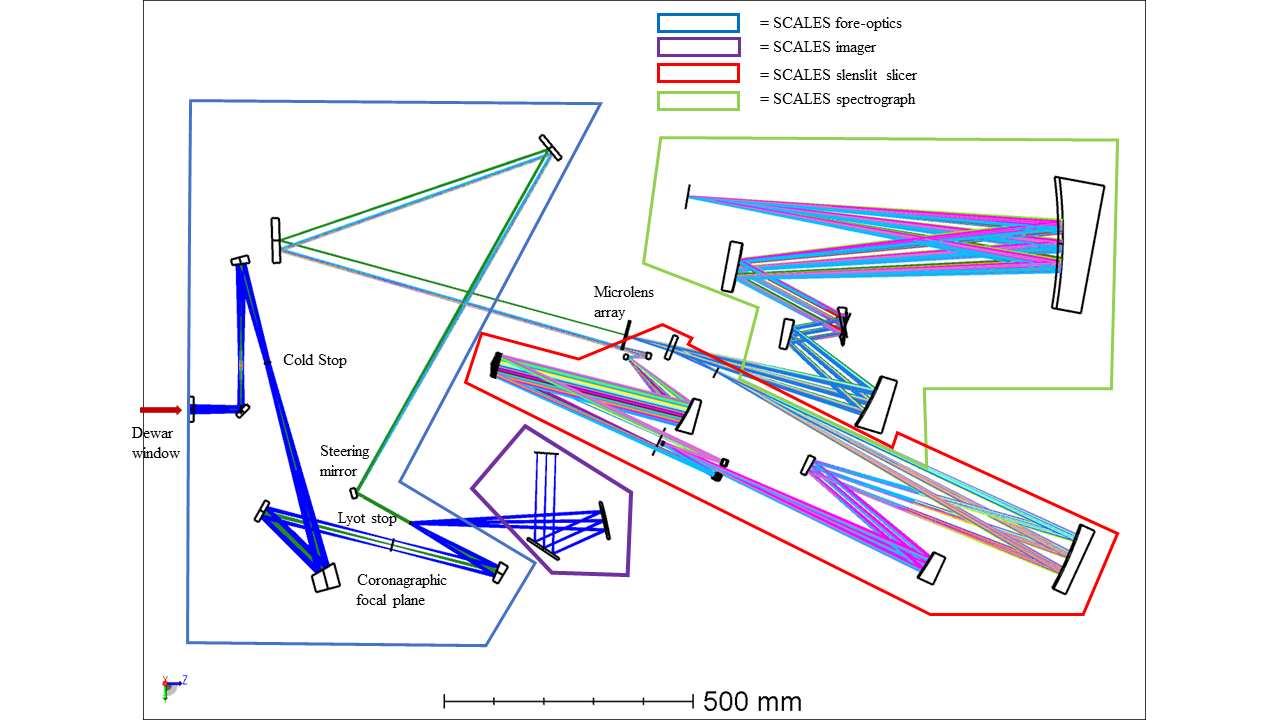}
    \caption{The full SCALES optical design, with subsystems foreoptics, imager, spectrograph and slenslit reformatter identified in colored boxes}
    \label{fig:FullSCALES}
\end{figure}

The optical design of SCALES is divided into four optical subsystems (Figure~\ref{fig:FullSCALES}): the fore-optics which host the coronagraphic optics and magnify the beam for sampling at the microlens array; the imager with a fixed plate scale, and the spectrograph which disperses the pupil images formed by the microlens IFU and records them on the spectrograph detector. The spectrograph itself is divided further into low and medium spectral resolution subsystems. 
\subsection{Keck Adaptive Optics}
\begin{figure}
    \centering
    \includegraphics[width=0.9\textwidth]{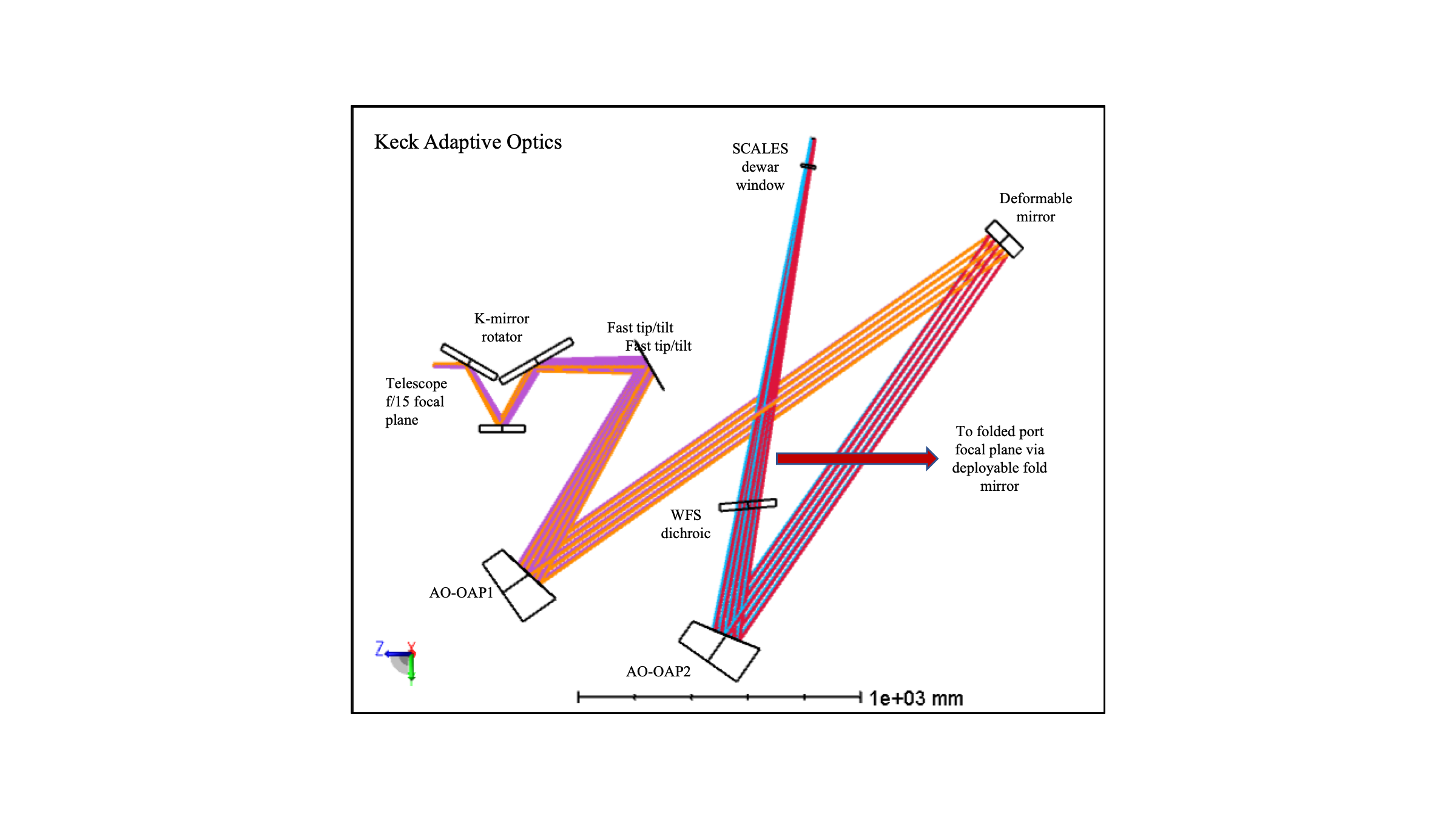}
    \caption{Optical path of the Keck adaptive optics system, located on the Nasmyth platform and serving as input to SCALES.}
    \label{fig:KeckAO}
\end{figure}

While not strictly a part of the SCALES instrument, the Keck AO system shown in Figure~\ref{fig:KeckAO} impacts the SCALES optical design and operation. 

The first optical component, near telescope Nasmyth focus, is a K-mirror de-rotator which can be used in either field-tracking mode to counteract the field rotation of the alt-az Keck (SCALES imager operation), or in pupil-tracking mode where it stabilizes the rotation angle of the pupil and allows the field to rotate (SCALES IFS operation).
The flat mirror following the K-mirror is a fast tip-tilt mirror used for field stabilization. It should be noted that the fast tip-tilt mirror does not reside at the pupil plane and introduces some pupil movement during operation. A set of matched off-axis parabolas (OAPs) re-image the pupil onto a deformable mirror (DM) and re-image the DM-corrected image plane into SCALES. A wavefront sensor dichroic in the converging beam allows shorter wavelengths to be diverted to the Shack-Hartmann wavefront sensors, using either natural or laser guide stars. A second dichroic can be utilized to send J or H-band natural guide star light to an infrared pyramid wavefront sensor\cite{BondPWFS2020}. Space near the K-mirror allows deployment of calibration sources, including white light and arc lamps. A calibration unit specific to SCALES at mid-IR wavelengths is under development and will be deployed at that location.

\subsection{SCALES foreoptics and Imager}
  \begin{figure}
    \centering
    \includegraphics[width=0.9\textwidth]{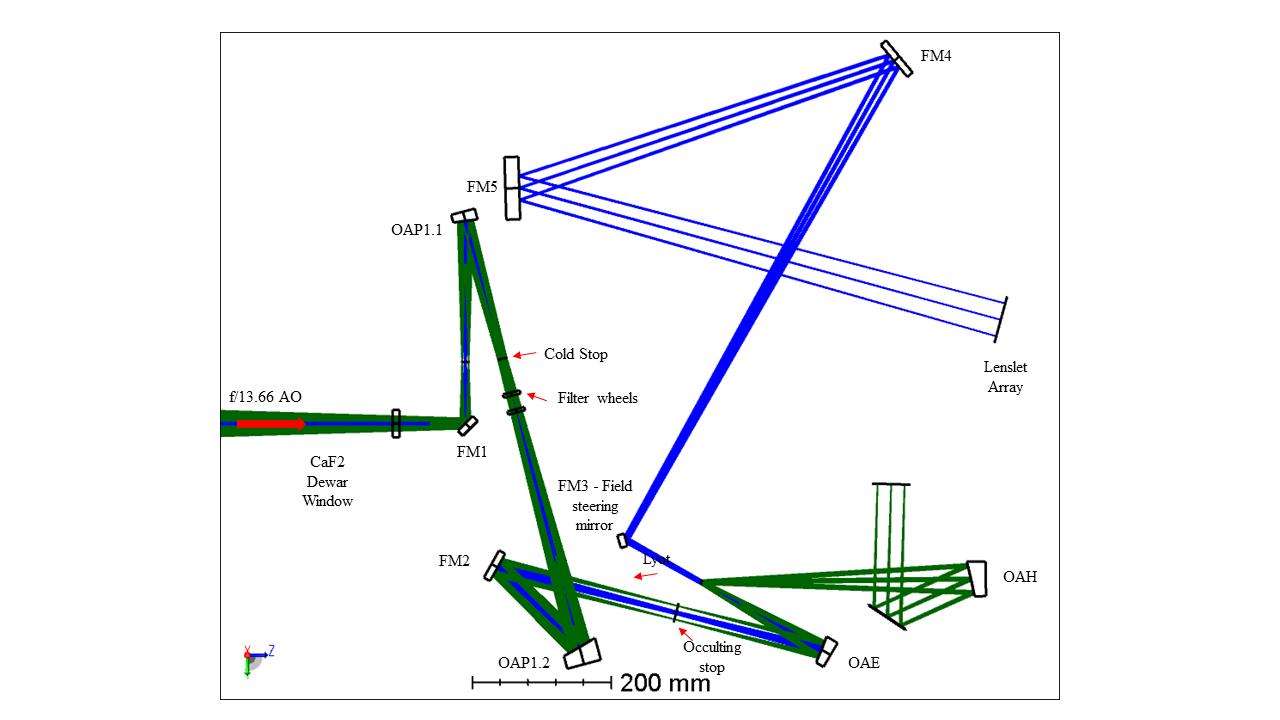}
    \caption{Detail of the SCALES foreoptics and imager optical designs.}
    \label{fig:Foreoptics}
\end{figure}
 The mid-infrared bandpass requires, as much as possible, the optics and detectors of SCALES to be housed in a cryogenic environment to minimize thermal background. Light enters the cryostat via a well-baffled, AR-coated calcium flouride window.  The foreoptics of SCALES serves many purposes: It must provide a well-formed pupil image for a cold-stop that is shared by coronagraph, imager and spectrograph and a collimated space for imager bandpass filters; for coronagraphy it must provide a focal plane for an occulting stop after the cold stop, and then a second well-formed pupil image for the Lyot stop. We require a space envelope for the imager sub-system, fed by reflected light from the Lyot stop position, and an unvignetted field of view of at least 12.3{\arcsecond} $\times$ 12.3{\arcsecond} to the imager pick-off location. Finally, the foreoptics must provide an unvignetted field of view of at least 3\arcsecond $\times$ 3\arcsecond to the IFS microlens location with $<$50 nm RMS wavefront error across the field of view at a plate scale of 0.02{\arcsecond} per spaxel (where a spaxel is equivalent to a single microlens) and enable the ability to steer the field between low- and mid-spectral resolution IFUs via a cryogenic steering mirror.  

 To achieve the above listed requirements we've implemented a pair of relays: The first consists of a 4-f relay with off-axis parabolas as powered optics, and the second consists of an off-axis ellipse which produces both the real pupil for the Lyot stop and the reformatted plate scale for the IFS. The cryogenic steering mirror resides just after the Lyot stop, close to the pupil to minimize pupil shift during field steering. The optical layout of just the fore-optics is shown in Figure~\ref{fig:Foreoptics}. 

There are four focal planes within the foreoptics: the f/13.66 Keck AO focal plane, an f/27 focal plane at the coronagraphic mask, an f/54 focal plane at the imager detector and an f/320 focal plane at the microlens array IFU. There are two pupil planes: the location of the cold stop and the coronagraphic Lyot stop.
\subsubsection{Imager}
\begin{figure}
    \centering
    \includegraphics[width=0.9\textwidth]{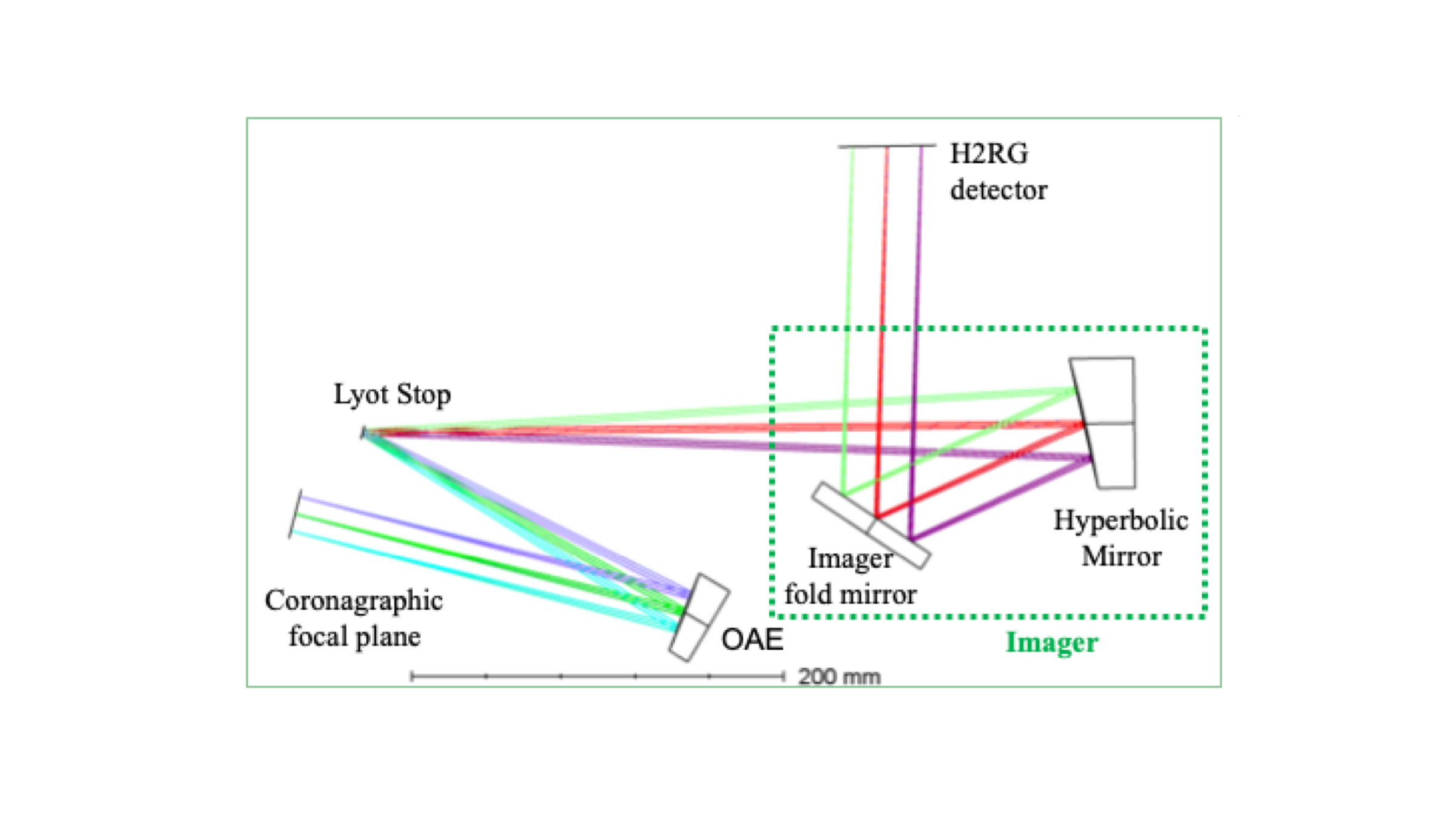}
    \caption{Detail of the SCALES imager optical designs.}
    \label{fig:ImagerOnly}
\end{figure}

The SCALES imager design is described in detail in a separate publication\cite{Banyal2022} but will be summarized briefly here. The Lyot wheel provides an ideal location for a pick-off mirror to the imager. The optical layout of the imager is shown in Figure~\ref{fig:ImagerOnly}. As seen in Table~\ref{tab:SCALESReqSum} the imager requires a 0.06\arcsecond per pixel plate scale over a 2k x 2k pixel$^2$ H2RG. Science requirements demand diffraction-limited imaging down to 1.0 $\mu$m wavelength and low grid distortion. Note that the imager is not a 4-f relay, as the off-axis ellipse in the foreoptics produces a slowly converging beam. Optimization therefore suggested a slightly hyperbolic off-axis asphere as the ideal camera optic for the imager, with a slightly tilted focal plane.  A fold mirror directs the f/54 beam to an area of the bench with ample space for the H2RG detector package, sidecar and associated baffling. 

\subsection{Foreoptics baseline performance}
All of the performance results for the foreoptics include contributions from the Keck telescope and adaptive optics system. The shape of the deformable mirror (DM) in the Zemax optical model (a "Zernike fringe surface") has been optimized to give best performance at the adaptive optics focal plane. Unless otherwise noted, results are monochromatic. In operation both telescope focus and DM shape will be adjusted depending on bandpass. The only refractive optics contributing to the wavelength dependence of wavefront quality are the wavefront sensor dichroic and cryostat window.

\subsubsection{Pupil quality}
\begin{figure}
    \centering
    \includegraphics[width=0.7\textwidth]{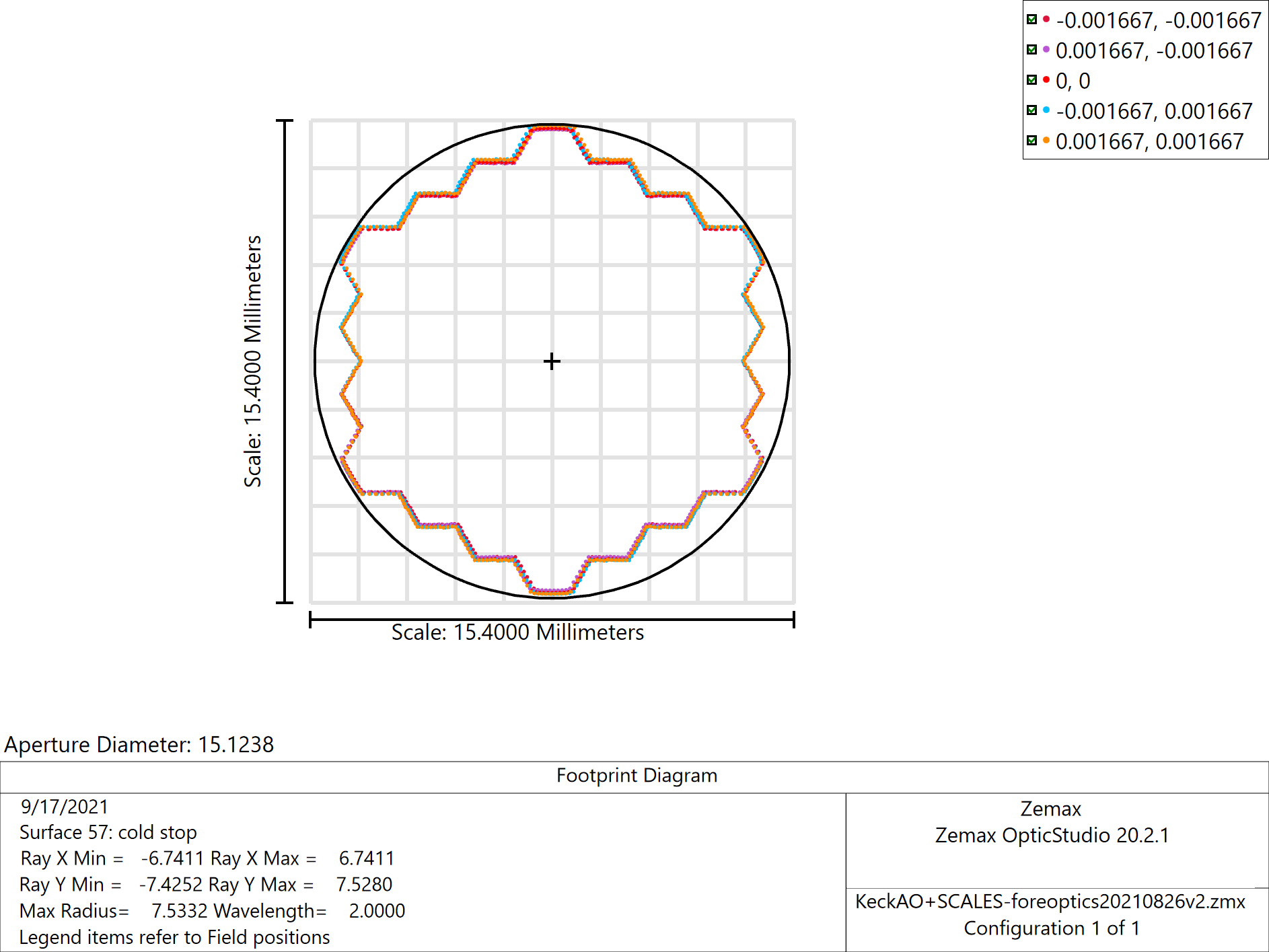}
    \caption{Beam footprint at cold stop located at image of the telescope primary. Beam diameter is 15.12 mm, and beam footprints from center and extreme field points overlap to within 60 $\mu$m.}
    \label{fig:ColdStopFootprint}
\end{figure}

Having a high quality, unaberrated pupil image allows suppression of background at the cold stop and proper coronagraphic masking at the Lyot stop with minimal throughput loss. Figure~\ref{fig:ColdStopFootprint} shows the footprint of the pupil at the cold stop for on-axis and corners of a 13\arcsecond $\times$ 13\arcsecond rectangular field of view. This pupil is 15.12 mm in diameter. The pupil footprint from all field points overlap to within 60 $\mu$m. 

\begin{figure}
    \centering
    \includegraphics[width=0.6\textwidth]{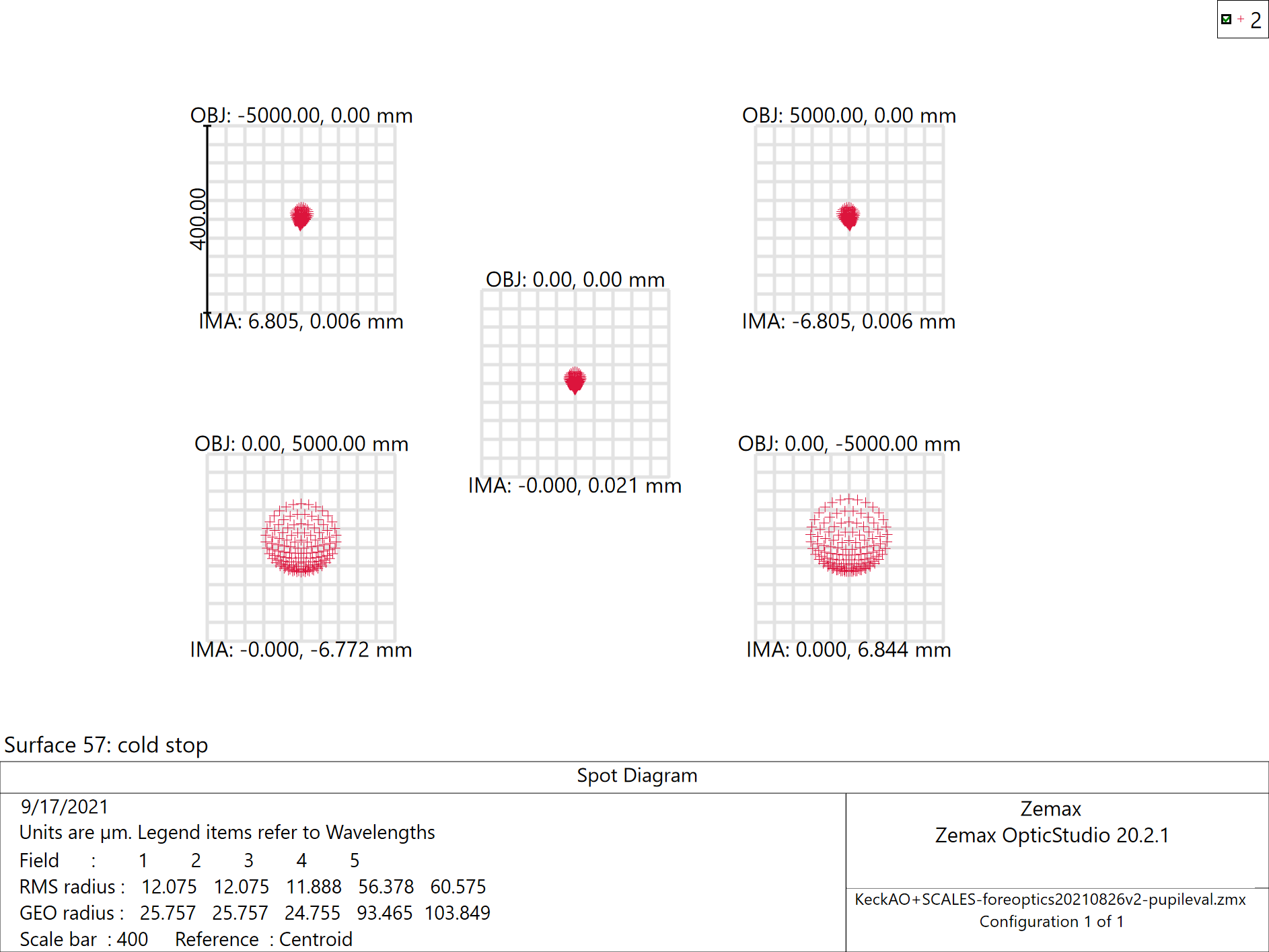}
    \caption{Evaluation of the pupil quality by geometric ray trace of primary mirror to cold stop. On-axis (center of primary) and pupil edges are shown. Pupil tilt (caused by OAP relay) causes a 60 $\mu$m blurring of the pupil edge over the imager field of view.}
    \label{fig:SpotsColdStopPupil}
\end{figure}

A second technique for evaluating pupil quality is a ray trace in which the object is the primary mirror and the field stop defines the aperture size. The resulting spot diagram allows evaluation of the blurring at the pupil edges. Figure~\ref{fig:SpotsColdStopPupil} illustrates the tilt of the pupil at the cold stop, causing blurring of up to 60 $\mu$m. This blurring represents 0.3\% of the primary diameter.
An identical analysis was carried out for the real pupil formed by the off-axis ellipse in the second relay, at the position of the Lyot stop. The low-resolution IFS field of view (2.2\arcsecond $\times$ 2.2\arcsecond) was used for this analysis. The pupil at the Lyot stop is 6.88 mm in diameter. Pupil quality given this small field of view is exceptional, with $\sim$4 $\mu$m blurring at the edges, or 0.06\% of the pupil diameter. 
\begin{figure}
    \centering
    \includegraphics[width=0.9\textwidth]{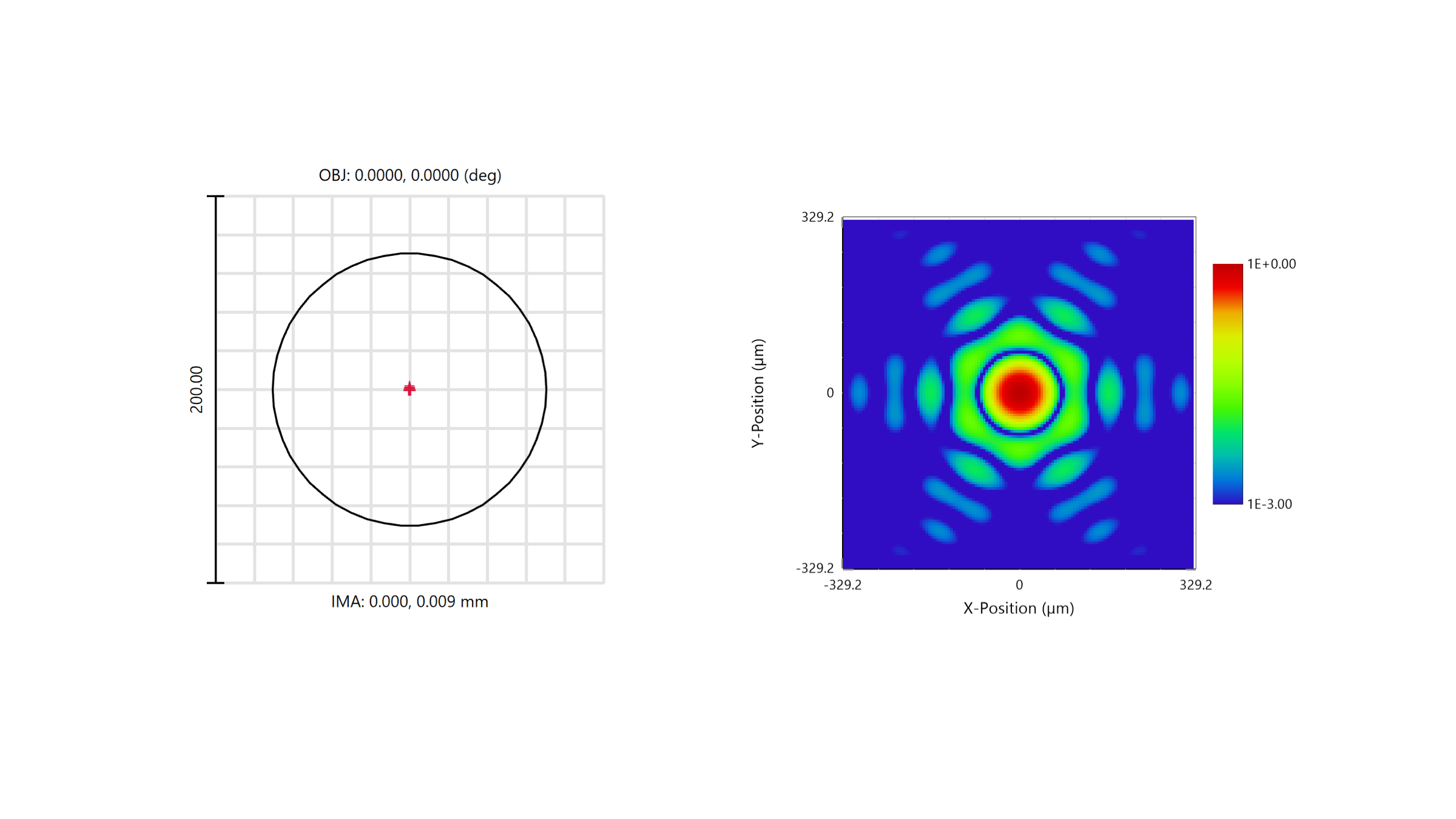}
    \caption{On-axis geometric spot diagram (left) and Huygens PSF at 2$\mu$m wavelength (right) at focus of first foreoptics relay. Circle on spot diagram at left indicates diffraction limit at 0.6 $\mu$m}.
    \label{fig:OccultingFocalPlane}
\end{figure}

\subsubsection{Performance of first relay}
Referring to Figure~\ref{fig:Foreoptics}, the OAP relay forms an intermediate focus via an f/27 converging beam which is folded to focus by a flat mirror. Located at this intermediate focal plane is the occulting mask of the coronagraph described in Section~\ref{sec:CoronagraphFocalPlane}. The performance of the coronagraph in terms of contrast is intricately related to the optical performance of this first relay. Figure~\ref{fig:OccultingFocalPlane} shows the on-axis spot diagram and predicted Huygens point spread function. These are consistent with the predicted 14 nm maximum wavefront error, the performance one might expect from a series of OAP relays used on-axis. 

\subsubsection{Imager performance}
\begin{figure}
    \centering
    \includegraphics[width=0.9\textwidth]{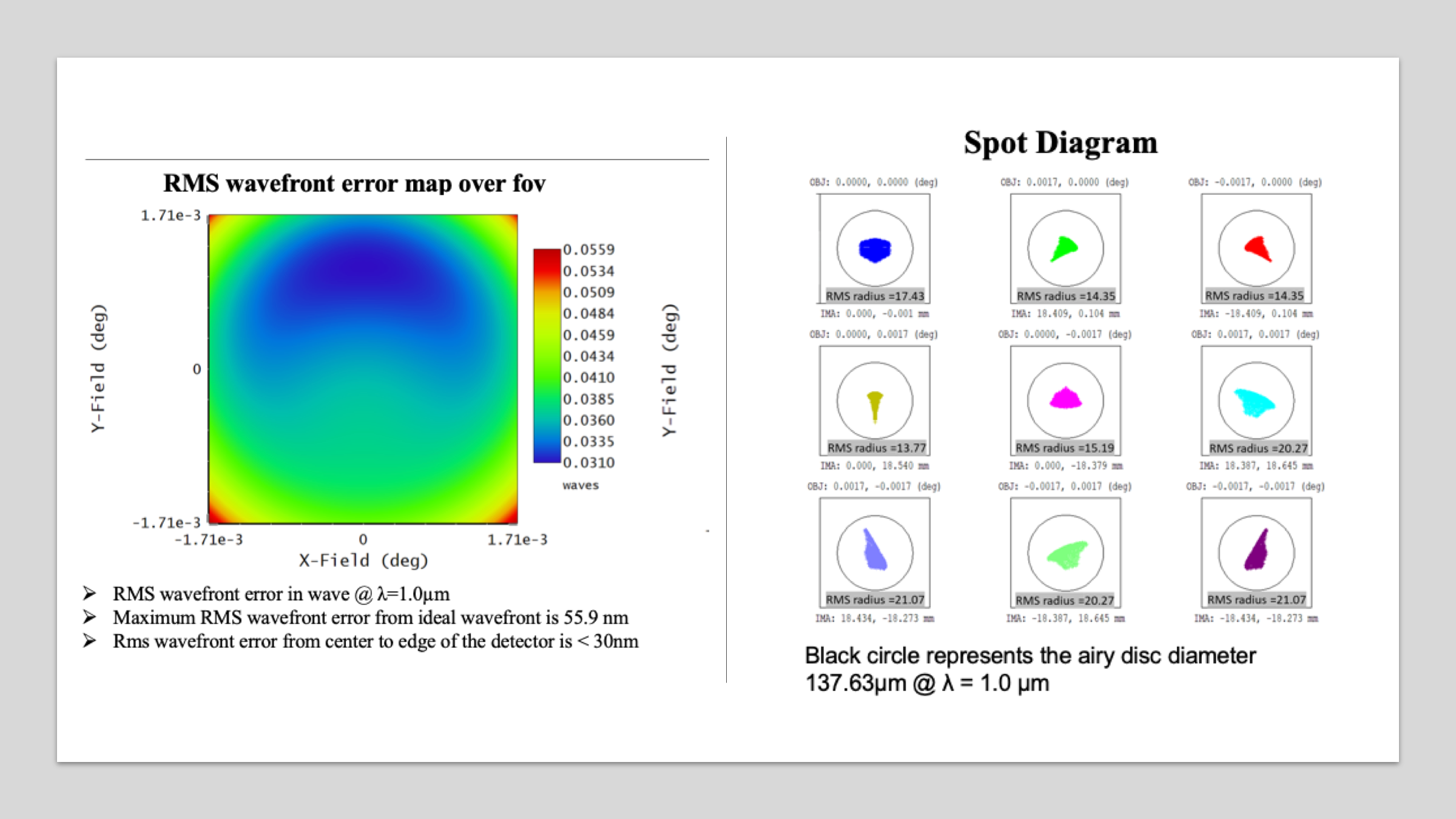}
    \caption{Imager performance in terms of diffraction-limited spot diagram at 1.0 $\mu$m wavelength (right) and RMS wavefront error less than 60 nm across the full 12.3\arcsecond $\times$ 12.3\arcsecond field of view (left)}
    \label{fig:ImagerPerf}
\end{figure}

As an adaptive optics-fed imager, the requirements for non-correctable (field-dependent) wavefront error are very stringent at less than 65 nm RMS wavefront error across the full field of view. The design meets this requirement with additional allocation for fabrication and misalignment tolerances. It also exhibits less than 1\% grid distortion over the full field of view and is highly telecentric (to 0.05 degree) at the detector to minimize plate scale changes with defocus. Full field spot diagram and RMS wavefront error can be seen in Figure~\ref{fig:ImagerPerf}. Additional results can be seen in Banyal\cite{Banyal2022}.
\subsection{Performance at microlens array integral field unit}
The second foreoptics relay consists of an off-axis ellipse and three fold mirrors to magnify and direct the beam to the position of the lenslet array IFU. The off-axis ellipse forms a real pupil 203 mm after the optic. A cryogenic tip/tilt mirror located shortly after the pupil allows field steering between spectrograph modes at the IFU, which resides at a magnified focal plane with a focal ratio of f/320 more than two meters downstream. 

The lenslet array defines the plane at which the spatial sampling of the integral field spectrograph takes place. Very low wavefront error and optical distortion are required at this focal plane to preserve the spatial resolution afforded by the adaptive optical correction, and allow precise astrometric positioning on the lenslet array. 
\begin{figure}
    \centering
    \includegraphics[width=0.9\textwidth]{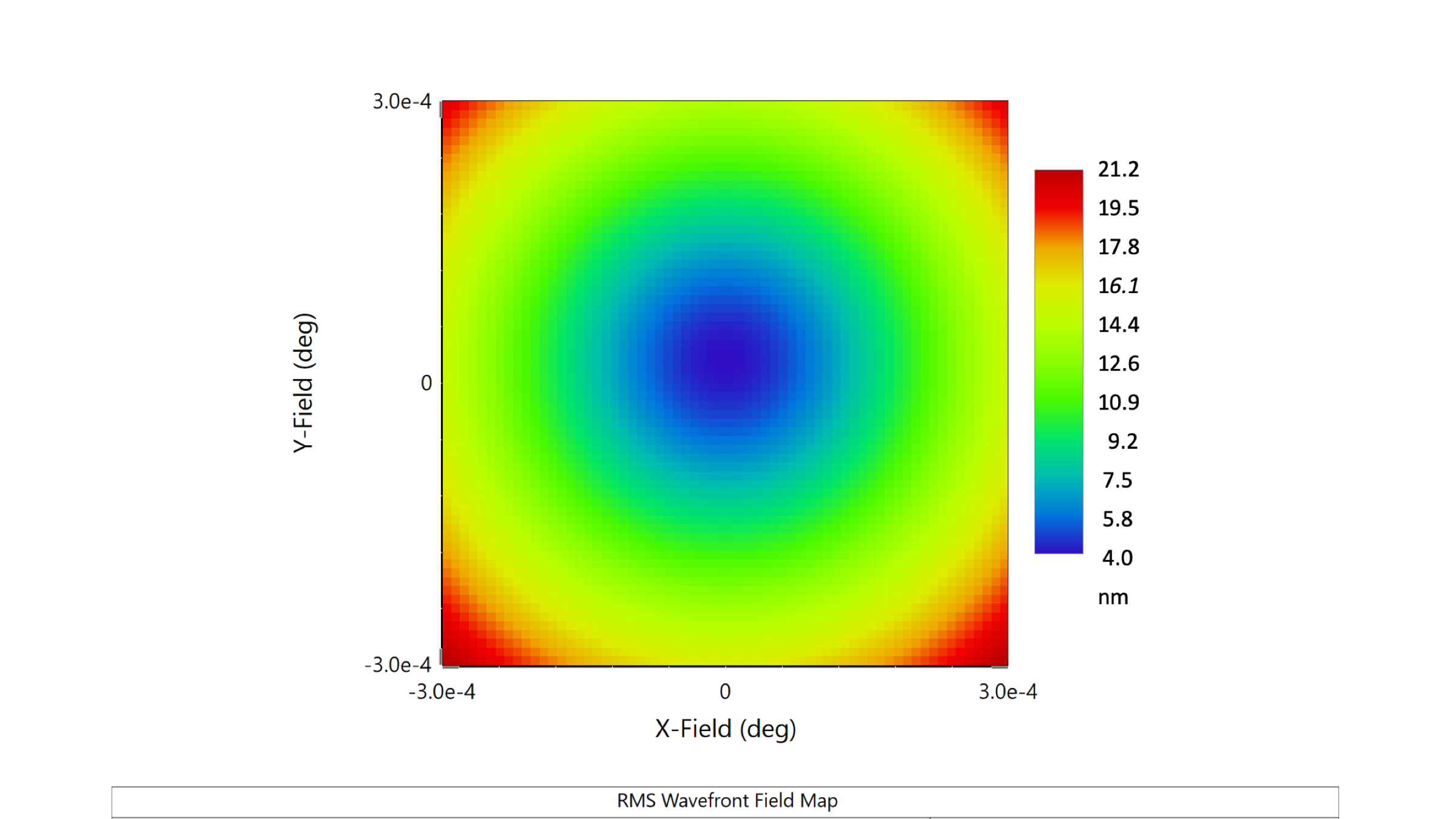}
    \caption{As-designed RMS wavefront error vs. field position at the integral field spectrograph input focal plane (microlens array IFU). Field of view shown corresponds to 108 $\times$ 108 $\times$ 341 $\mu$m pitch lenslets, or 2.2\arcsecond $\times$ 2.2\arcsecond}.
    \label{fig:WFElensletIFU}
\end{figure}
Figure~\ref{fig:WFElensletIFU} shows the variation of RMS wavefront error with field of view over the full 108 $\times$ 108 spaxels$^2$ of the lenslet array which is $<$20 nm over nearly the full 2.2\arcsecond $\times$ 2.2\arcsecond field of view. The maximum grid distortion at the lenslet array is only 0.1\%.

\subsection{SCALES pupils and coronagraph architecture}
\subsubsection{Cold and Lyot stops}
The telescope exit pupil is re-imaged onto the SCALES cold stop pupil plane where a pupil-shaped aperture is placed to suppress stray light and thermal background. During IFS operations we will coordinate with the AO K-mirror to keep the pupil rotation angle fixed: the field rotation is actually helpful in direct imaging of exoplanets and allows Angular Differential Image post-processing. During imager operations the field is de-rotated with the K-mirror, and the cold stop rotates within its mount to match the rotation caused by the alt-az telescope mount. 

\begin{figure}
    \centering
    \includegraphics[width=0.9\textwidth]{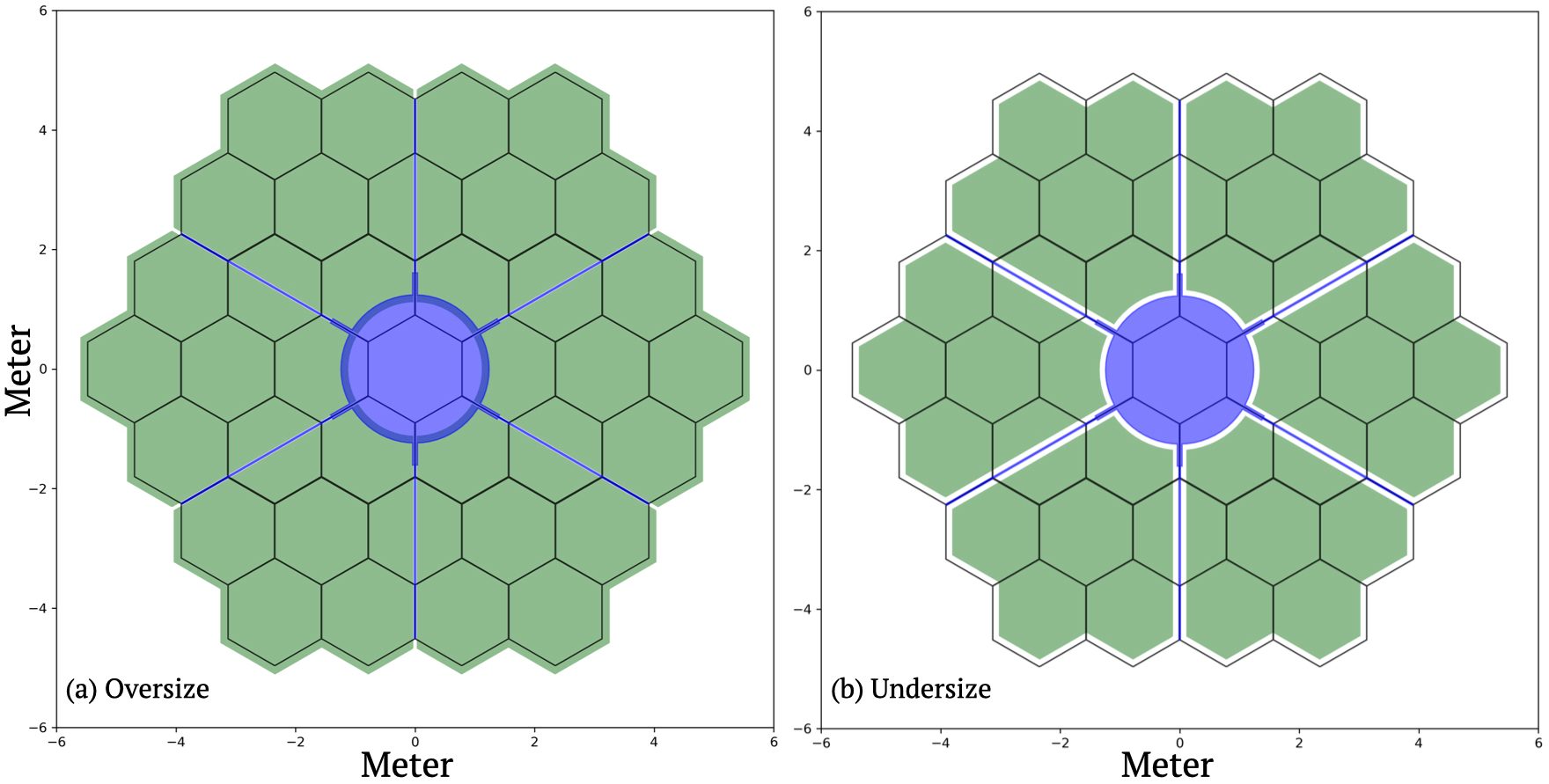}
    \caption{The left panel shows the SCALES cold stop design and the right panel shows the SCALES Lyot stop design. Transmitted regions are shown in green. The cold stop lets the entire clear aperture of the primary throught, accounting for 2\% pupil nutation. The Lyot stop ensures a constant pupil geometry, also accounting for the 2\% nutation.}
    \label{fig:ColdStopLyotStopDesign}
\end{figure}

SCALES uses the common approach of a focal plane coronagraph to reject on-axis starlight and a pupil plane Lyot stop to reject diffracted starlight. We summarize work by Li\cite{Jialin2021,Jialin2022} in design of the cold and Lyot stops (Figure~\ref{fig:ColdStopLyotStopDesign}). The cold stop will be slightly oversized; we plan to use a stop that allows light from the full Keck primary to pass, assuming 2\% pupil nutation (from K-mirror misalignment). The Lyot stop design is undersized and includes the diffractive effects of the coronagraph. The Lyot stop wheel has room for several stops, so we  plan to purchase several with varying outer diameters to optimize on-sky performance. 

\subsubsection{Coronagraphic focal plane}
\label{sec:CoronagraphFocalPlane}
The coronagraphic focal plane has a plate scale of 0.65 \arcsecond/mm. Two vector vortex coronagraphs that are currently installed in NIRC2 will be moved to SCALES as the baseline coronagraphic mode, courtesy of Dmitri Mawel and Olivier Absil\cite{Absil2016}. The focal plane masks are mounted in a linear stage to allow selection between K-band and L- and M-bands, as well as an open position (no coronagraph) and a pupil-imaging lens. 

\subsection{Lenslet Array}
Table~\ref{tab:lenslet-array-geometric-reqs} states the geometric requirements adopted for the microlens array IFU. 
\begin{table}[hpt]
    \centering
    \caption[Summary of lenslet array geometric requirements.]{
        Summary of the geometric requirements of the lenslet array.
    }

\begin{tabular}{p{5cm}  p{8cm}}
\hline
Low-Resolution Lenslets & $108 \times 108$ (or larger to enable alignment) \\
\hline
Medium-Resolution Lenslets & $17 \times 18$ \\
\hline
Lenslet Size & $341 \times 341$ {$\mu$m} \\
\hline
Lenslet Focal Ratio & f/8 \\
\hline
\end{tabular}

    \label{tab:lenslet-array-geometric-reqs}
\end{table}
The low- and medium- resolution lenslet array will be on a single substrate, with identical parameters to simplify mounting and alignment. The lenslet array will be fabricated from silicon for ease of manufacture and good transmission out to 5.0 $\mu$m. 

\subsubsection{Lenslet diffraction and pinholes}
Diffraction at the microlens pupil grid is important to characterize for two reasons: Diffraction has the effect of shifting the best focus away from geometric focus and diffraction from the square lenslet aperture can cause crosstalk between adjacent spectra in lenslet-based integral field spectrographs. The crosstalk can be mitigated by a pinhole mask located at the pupil grid plane. To find the best location for the pinhole mask, and the optimal size for the pinholes we modeled the lenslet array and spectrograph in physical optics. 

\begin{figure}[htp]
    \centering
    \begin{minipage}[c]{0.31\textwidth}
        \includegraphics[width=0.98\textwidth]{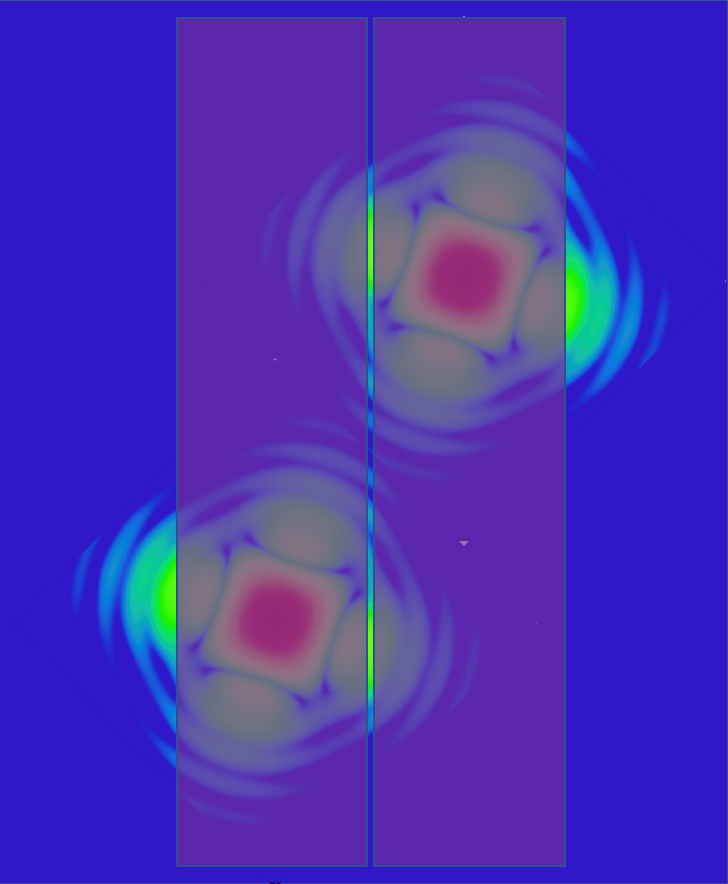}
    \end{minipage}
    ~
    \begin{minipage}[c]{0.65\textwidth}
        \includegraphics[width=0.98\textwidth]{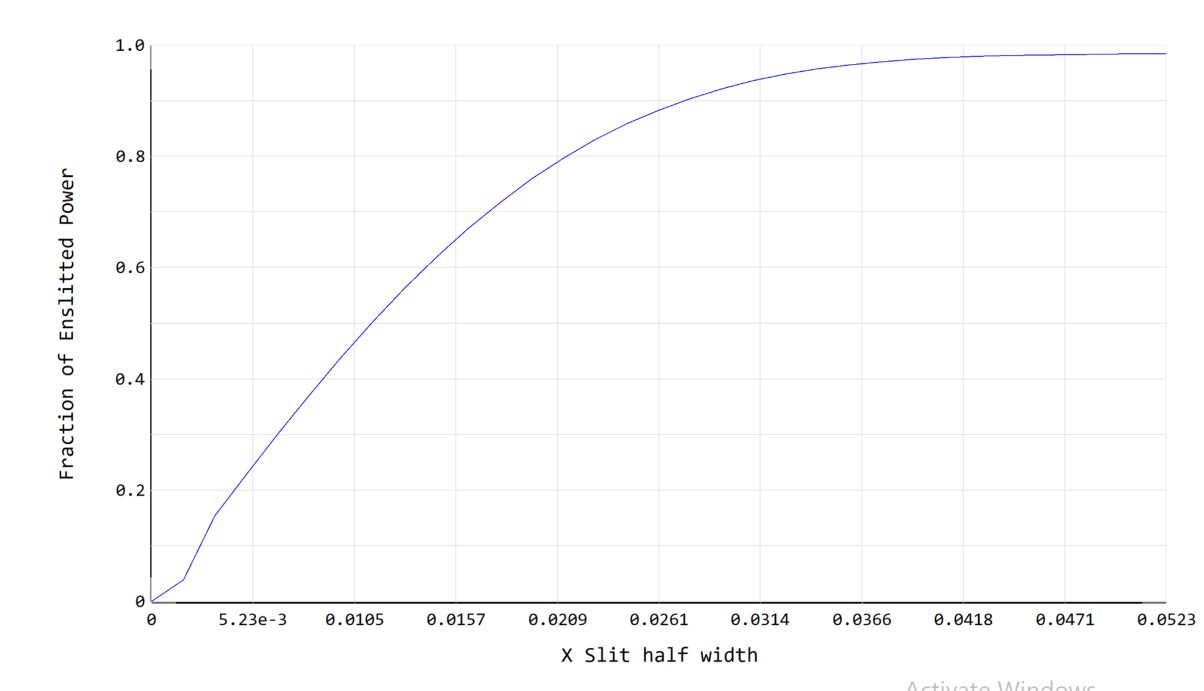}
    \end{minipage}
    \caption[2D and 1D representations of cross-talk.]{
         Left: Log-scale images of two lenslet PSFs, with the direction of the spectral traces super-imposed.  
         Light that falls outside of the semi-transparent rectangles becomes crosstalk in the adjacent spectra.  
         For the purposes of calculating crosstalk, we use the Zemax enslitted energy tool and use 108 $\mu$m (the separation between adjacent spectra) as our figure of merit. 
         Right: Enslitted energy of an individual lenslet+pinhole through the SCALES spectrograph at $\lambda=$ 5 $\mu$m.  
         Light that falls outside of a slit width of 108 µm (the distance between adjacent spectra) becomes crosstalk.  
         The x-axis of the plot is the slit half-width.  
         Therefore, we are interested in how much light falls outside of a half-width of 54 $\mu$m.  
         The output suggests $\sim 2$\% crosstalk.  
    }
    \label{fig:enslitted-energy}
\end{figure}

Diffraction for our system was modeled using the Zemax physical optics package. The 341 $\mu$m $\times$ 341 $\mu$m pitch lenslet with a focal length of 3.86 mm produces an f/8 beam. Each lenslet forms a pupil image with a geometric diameter of 8 $\mu$m at approximately one focal length behind the lenslet array. The diffraction limit for an f/8 beam varies from 16 - 40 $\mu$m over the 2.0 - 5.0 $\mu$m bandpass, so we can approximate the pupil image as a diffraction-limited point source which is reimaged onto the detector with a magnification of 1.0. 
\begin{figure}
    \centering
    \includegraphics[width=0.9\textwidth]{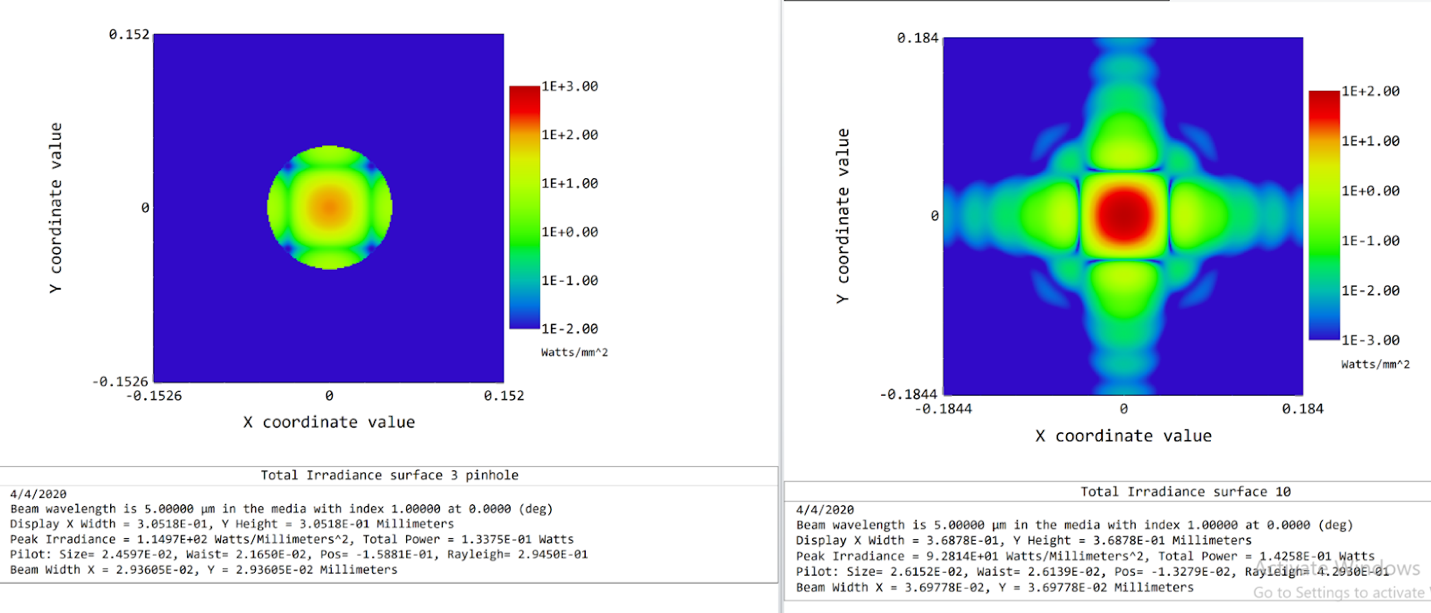}
    \caption{Physical optics propagation model for a single lenslet of the microlens IFU with a pinhole mask at the micro-pupil (left) and the image of that micro-pupil and pinhole mask at the detector after propagation through the SCALES spectrograph.}
    \label{fig:lenslet-pointsource-thru-pinhole}
\end{figure}
We position a circular pinhole at 2.25 mm after the silicon lenslet back surface (distance of best focus at 3.5 $\mu$m wavelength, determined with physical optics propagation). Figure~\ref{fig:lenslet-pointsource-thru-pinhole} shows the f/8 micro-pupil image masked by a 108 $\mu$m pinhole and that same masked pinhole imaged through the spectrograph optics. With a 6-pixel spectral spacing, light diffracted beyond the 108 $\mu$m pinhole diameter will intersect with the adjacent spectra. A smaller pinhole will reduce crosstalk, but also impact throughput. To quantify the pinhole suppression of crosstalk, an enslitted energy analysis was performed. Light that falls outside of a slit width of 108 $\mu$m (the separation between adajcent spectral traces) becomes crosstalk, so enslitted energy within this diameter is our figure of merit. A plot of enslitted energy versus radius from pinhole center is shown in Figure~\ref{fig:enslitted-energy}. This analysis suggests a 2\% crosstalk at $\lambda = 5.0 \mu$m which is sufficient to our requirements. 

\subsection{Spectrograph Optical Design}
The spectrograph input is made up of the lenslet array's micro-pupil images at the pinhole mask. It's a regular grid of 341 $\mu$m spacing with a uniform beam speed of f/8. The low spectral resolution grid contains 108 $\times$ 108 spaxels, while the medium spectral resolution samples a smaller field at 17 $\times$ 18 spaxels (see \ref{tab:lenslet-array-geometric-reqs}). 

\begin{figure}
    \centering
    \includegraphics[width=0.9\textwidth]{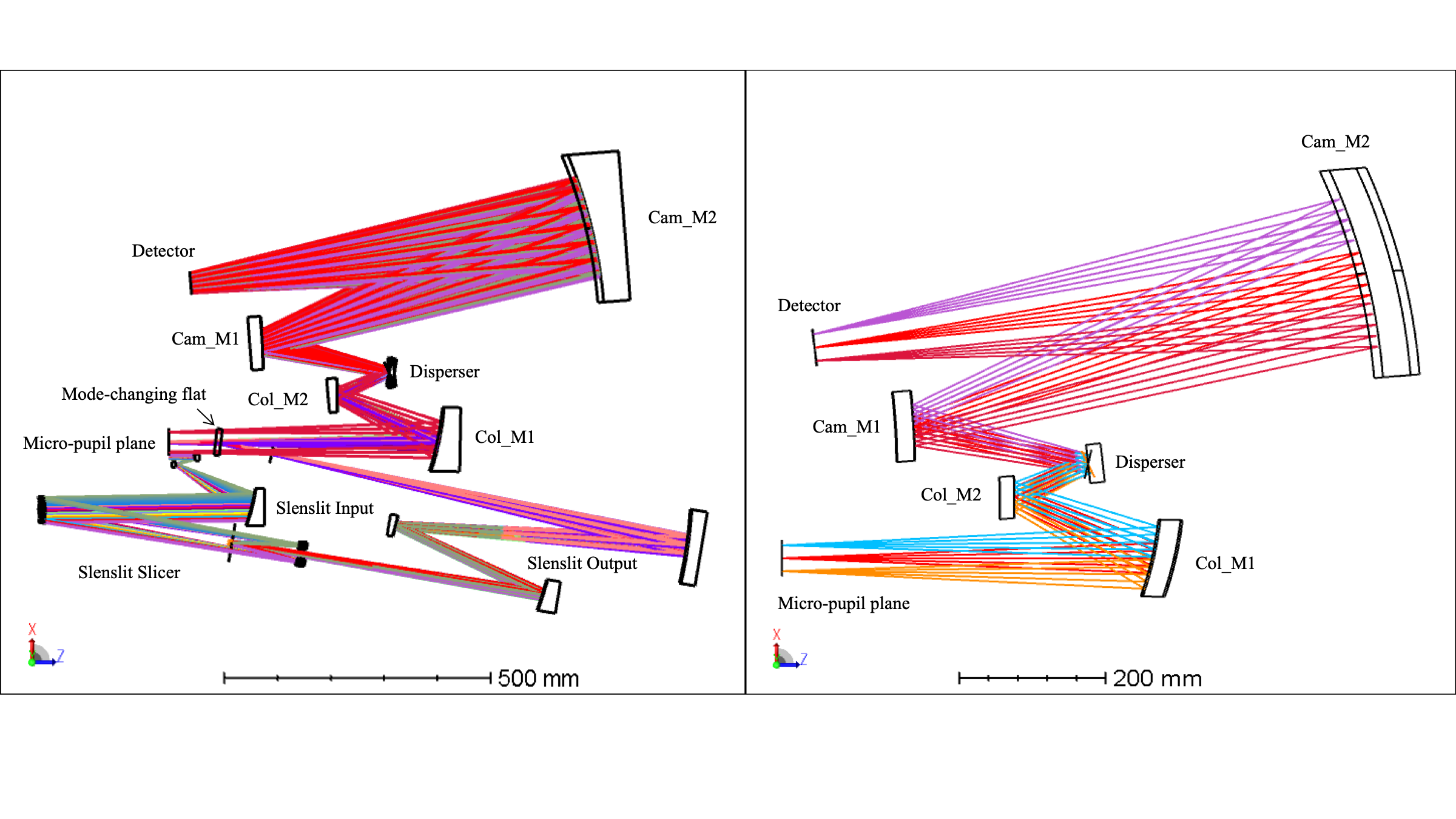}
    \caption{Spectrograph (right) and spectrograph plus slenslit (left) optical layouts. The micro-pupil grid formed by the lenslet array is input either directly into the spectrograph for low-resolution spectroscopy (right), or through a series of optical relays which reformats the micro-pupils into a pseudo-slit for medium resolution spectroscopy (left).}
    \label{fig:spectrographlayout}
\end{figure}

Both the low and medium spectral resolution modes utilize the 2-element f/8 collimator and camera, the elements of which are high order, off-axis aspheres. Prior to the collimator, the smaller pupil grid of the medium resolution mode is reformatted by a set of image slicing optics into a three row pseudo-slit, which is dispersed by a selection of gratings across the full length of the spectrograph detector. The low resolution mode utilizes double pass LiF prism dispersers, clocked to interleave the spectra for each spaxel across 60 pixels on the detector. The optical paths from pinhole grid to detector are shown in Figure ~\ref{fig:spectrographlayout} for both low- and medium-resolution modes. The detector is slightly tilted to improve optical performance and direct ghost reflections away from the optical path. 

\begin{figure}
    \centering
    \includegraphics[width=0.75\textwidth]{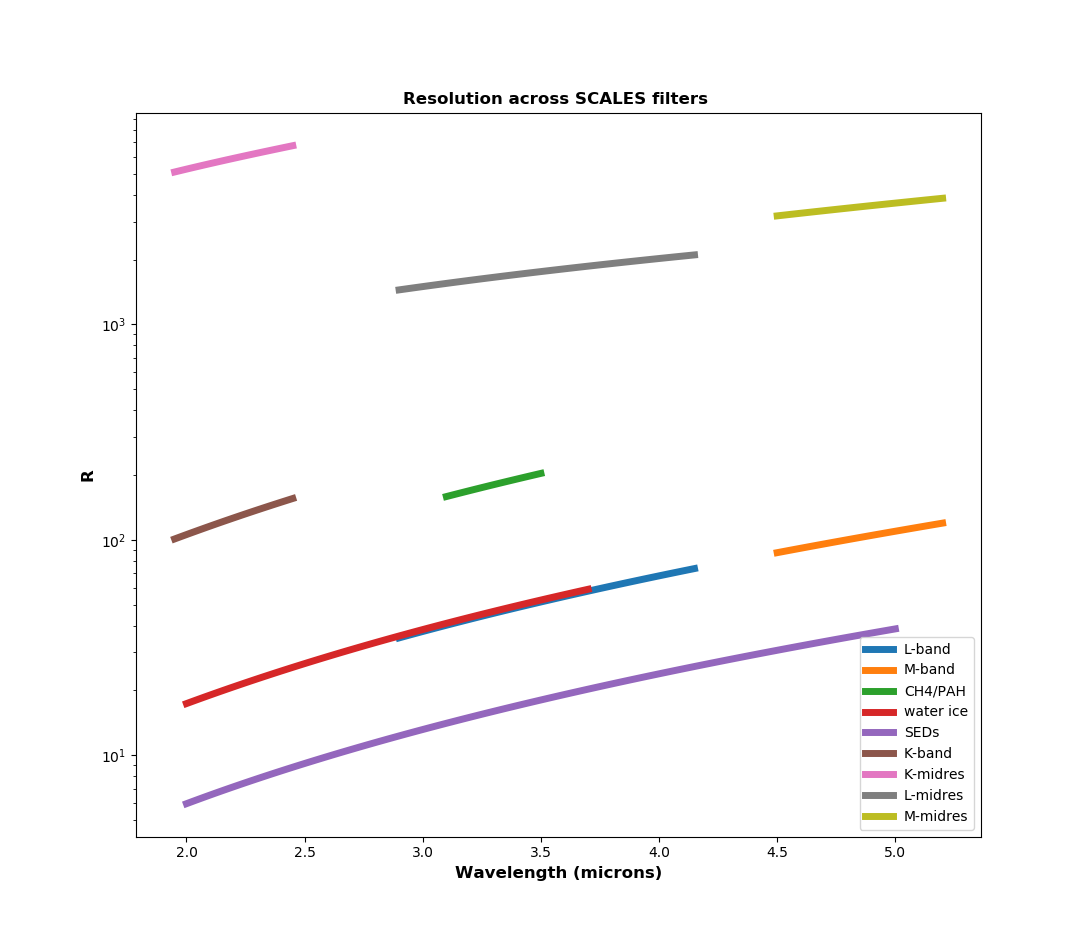}
    \caption{SCALES as designed spectral resolution.}
    \label{fig:AllModesSpecRes}
\end{figure}

Selection between the modes is accomplished with a flat mirror on a linear stage which folds in the light from the slenslit module while blocking the low-resolution lenslet grid, or alternatively blocks the slenslit path while in low-resolution mode. A disperser carousel holds a selection of six LiF prisms and three reflective ruled gratings, and these dispersers in combination with order blocking filters located near the detector provide all of the bandpass/resolution modes represented in Figure~\ref{fig:AllModesSpecRes}. The plot was created using the full spectrograph optical model, including prism and grating designs. 

The SCALES slenslit optical and mechanical design is described in detail in Stelter\cite{DenoSlenslit2021,DenoSlenslit2022}. Following the smaller 17 $\times$ 18 spaxel IFU is a second pinhole grid which serves as the object plane for an input relay to the slicer, converting the beam from the f/8 of the lenslet into the f/64 demanded by the slicer module. The slicer stack consists of 18 slices, one for each column of spaxels. The light is reformatted by the tilted, powered slicer mirrors in conjunction with a pupil and field mirror array. The powered, tilted field mirrors ensure a common entrance pupil to the slenslit output relay, which converts the beam back into the f/8 and co-locates the output focal plane, which now consists of 3 rows of pseudoslit, with the low-resolution pinhole grid. The beam is folded back into the spectrograph with the mode selection mirror, and is dispersed by one of three gratings in the disperser carousel.   

\subsubsection{Spectrograph performance}
The spectrograph must perform well enough to sample $\Delta\lambda$ at all passbands and resolutions at two pixels, or 36 $\mu$m. The baseline design goal for the spectrograph performance was therefore an RMS spot size $<$10 $\mu$m to allow signficant margin for fabrication and alignment errors. 

\begin{figure}
    \centering
    \includegraphics[width=0.8\textwidth]{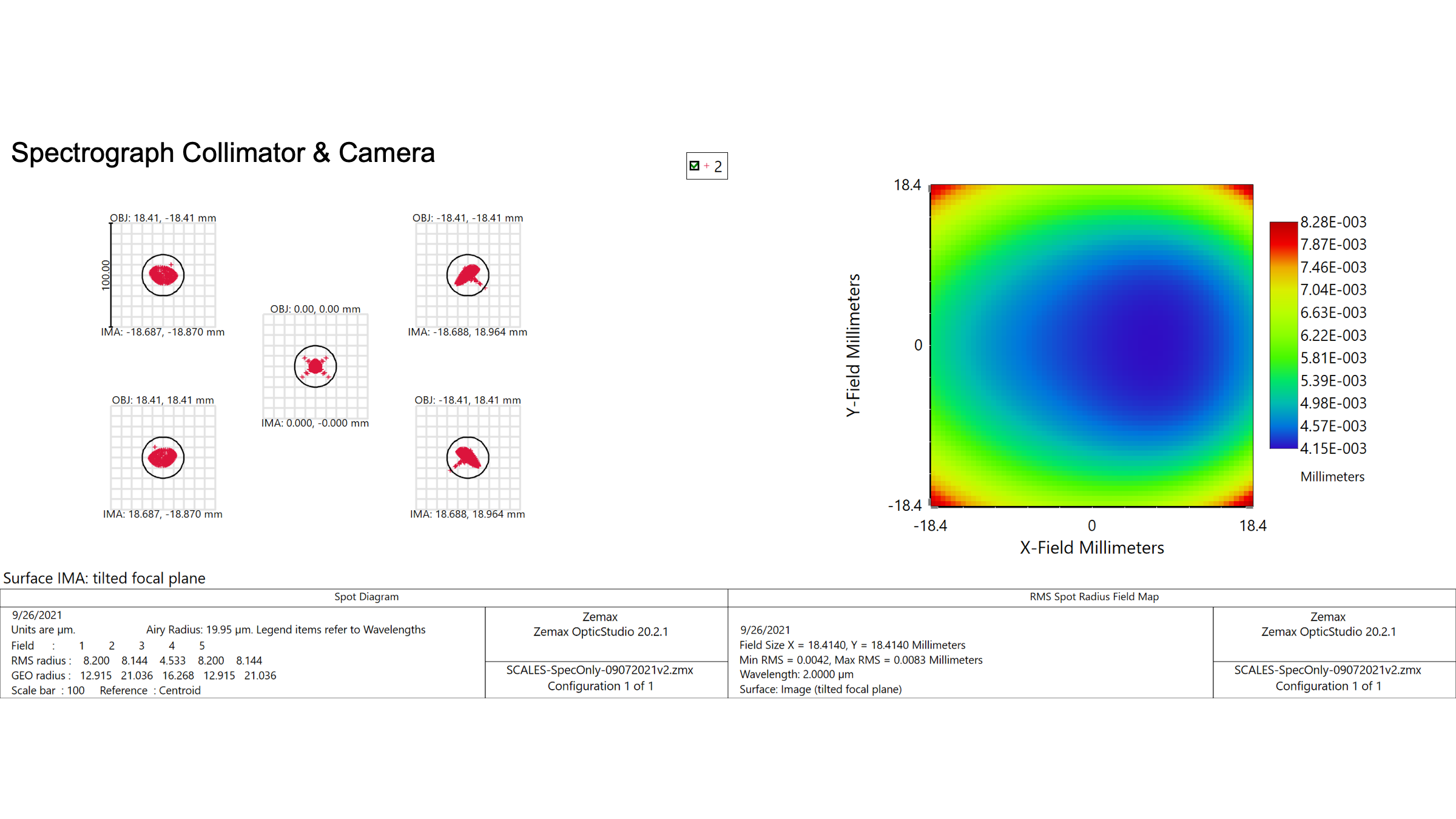}
    \caption{Baseline design spectrograph performance over the full low-resolution microlens array field of view. On-axis and corner spot diagrams are shown on the left, and full-field RMS spot radius over the full detector area are shown on the right. Circles in the spot diagram represent an Airy radius at 2 $\mu$m wavelength.}
    \label{fig:spectrographperf}
\end{figure}

Figure ~\ref{fig:spectrographperf} shows the spot diagram for on-axis and corners of the full low-resolution pupil pinhole grid, as well as a map of RMS spot radii over the full detector. This spot diagram considers only the spectrograph collimator and camera, and treats the pinhole grid as an array of point sources. It also includes only geometric aberrations (the circle in the spot diagram is the Airy radius at 2.0 $\mu$m wavelength. 

\begin{figure}
    \centering
    \includegraphics[width=0.8\textwidth]{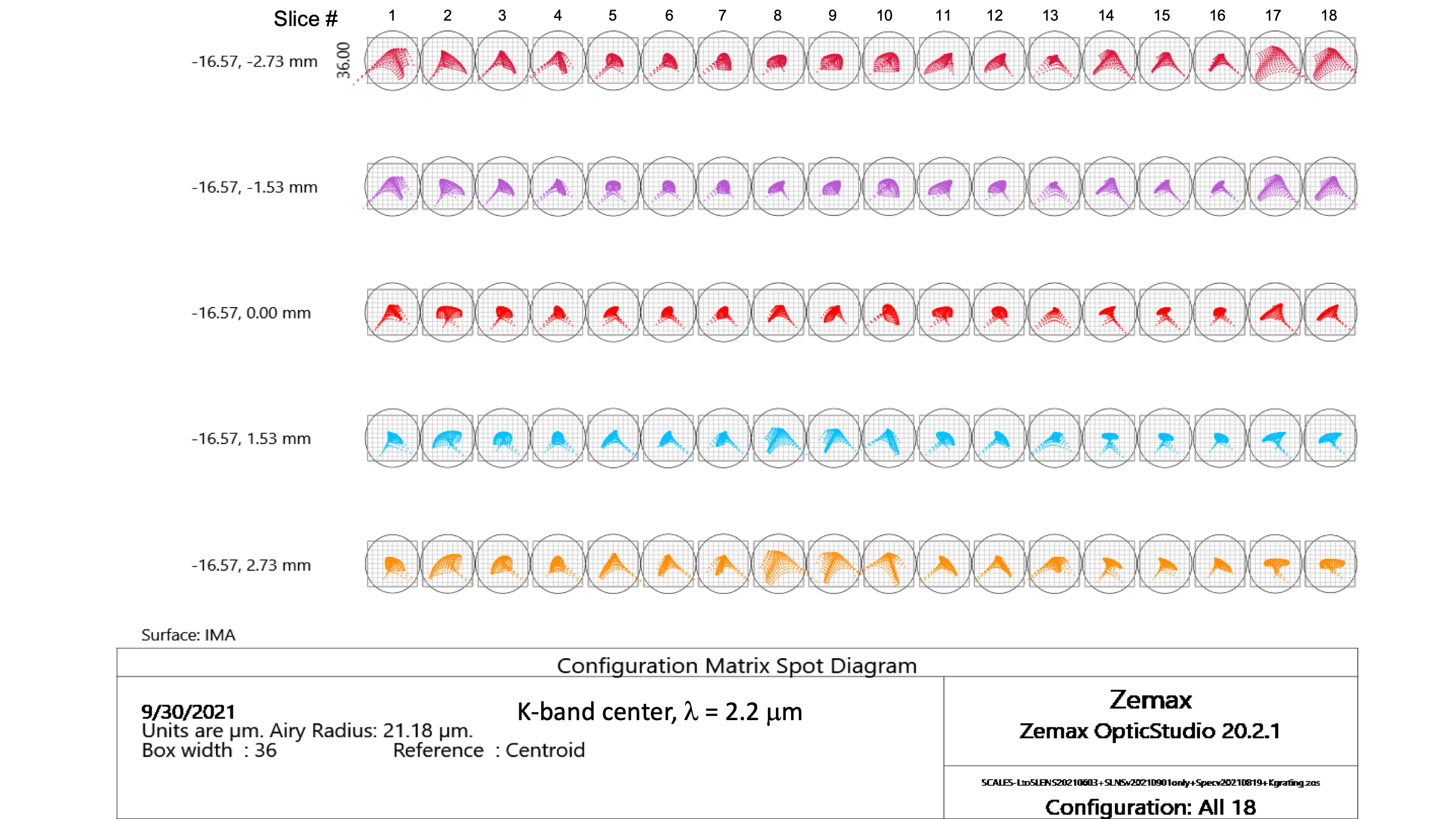}
    \caption{Baseline design slenslit spectrograph performance as spot diagrams for the mid-resolution field of view with the K-band grating, mid-band wavelength of 2.2 $\mu$m. Rows of spots are for five different spaxels along each of the 18 slices (columns).}
    \label{fig:medresConfigSpots}
\end{figure}

Although a smaller field of view, the medium resolution mode must maintain optical performance through the slenslit input relay, the slicer module, and the output relay as well as the spectrograph optics. The resulting pseudoslit fills the full width of the detector, and is dispersed across the full length. It therefore has the same field angle requirements on the camera as the low-resolution mode with its larger spatial field of view. Figure~\ref{fig:medresConfigSpots} shows a configuration spot matrix for the center K-band wavelength for five spaxel "field" points along each of the 18 slices. As with Figure~\ref{fig:spectrographperf} the circles represent the diffraction-limit Airy radius at K-band. Performance has been evaluated over the full wavelength range of the passband and found to be diffraction-limited for all medium resolution spaxels and for all passbands.

\section{OPTICAL MODELING}
\subsection{End-to-end performance}
\begin{figure}
    \centering
    \includegraphics[width=0.9\textwidth]{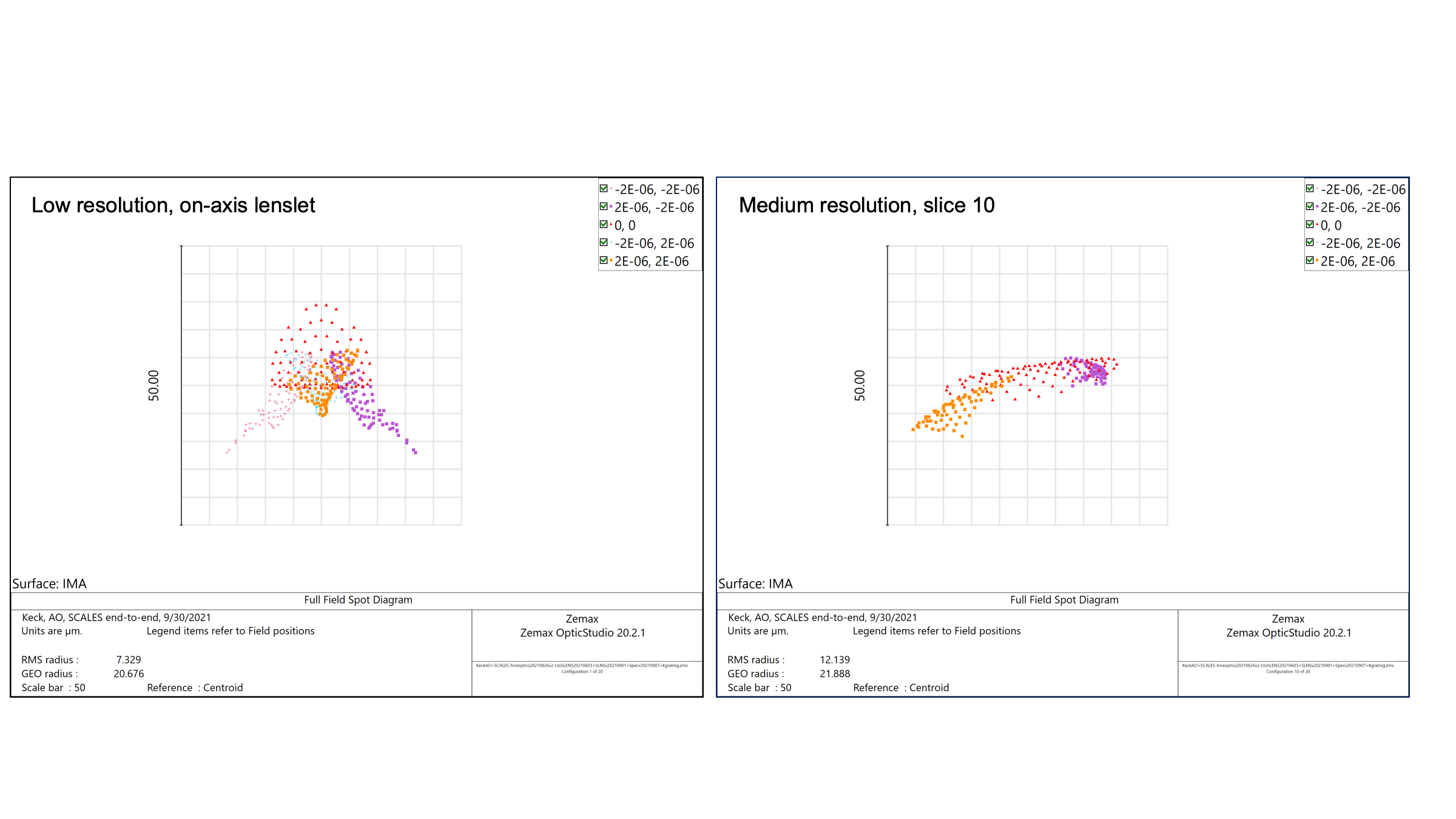}
    \caption{Full-field spot diagrams for a single spaxel traced from sky-to-detector for the low-resolution mode (left) and the medium resolution mode (right). Different colors represent different field points on a single spaxel. the box is 50 $\mu$m on a side.}
    \label{fig:etoespots}
\end{figure}

The SCALES team has used optical modeling derived from the Zemax optical design to inform the instrument simulator and data reduction pipeline\cite{Zack2020,Zack2021}, and to verify performance end-to-end. To model end-to-end performance, several field points are defined per spaxel. Recall that the lenslet array spatially samples the focal plane, and that each lenslet represents a 0.02\arcsecond $\times$ 0.02\arcsecond field of view. Rays are traced through the entirety of the optical system:  Keck telescope, adaptive optics system, SCALES foreoptics, through the lenslet array which forms a pupil image and on through the full spectrograph model. Performance can be checked via the full-field spot diagram in Zemax, which shows full spaxel spot diagrams. Examples of the full-field spot diagrams for both the low- and medium-resolution modes are shown in Figure~\ref{fig:etoespots}. Here different colors represent different points in a single spaxel. The combined field points for the low-resolution lenslet have an RMS radius of 9 $\mu$m, while the medium resolution combined spot from slice \# 10 is 12$\mu$m, both well under the diffraction limit at 2$\mu$m. 

\subsection{Spectral Layout}
\begin{figure}
    \centering
    \includegraphics[width=0.9\textwidth]{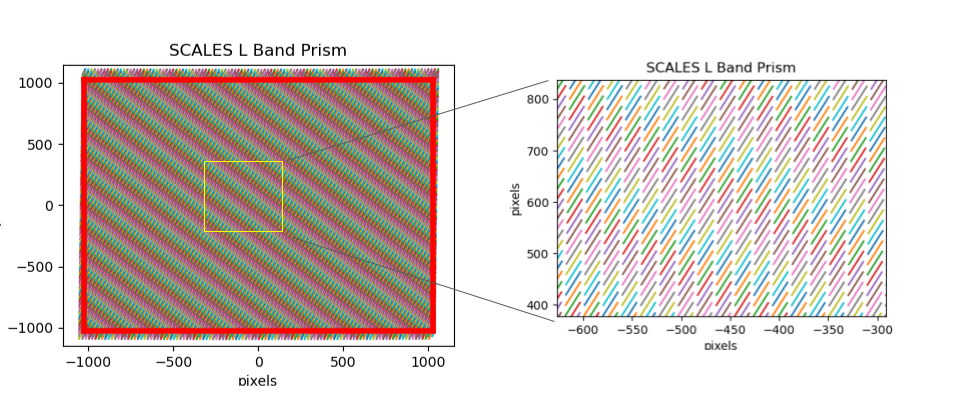}
    \caption{Spectral layout for the L-band low-resolution mode. Different colors are used to represent spectra from different spaxels.}
    \label{fig:LbandLowResSpec}
\end{figure}

To map the spectral layout, we define a field point at the center of each spaxel and use Zemax macro scripts to find the corresponding centroids on the detector for all spaxels with a good sampling of the wavelength dispersion. The spectral layouts for all prism and grating modes were checked for missing and partially missing spectra from edge spaxels, distance from adjacent spectra (to ascertain potential for crosstalk between spectra) and the dispersion solution at the detector to check the resolution achieved in different modes (Figure~\ref{fig:AllModesSpecRes}). An example of the prism spectra layout for the L-band passband is shown in Figure~\ref{fig:LbandLowResSpec}. 

\begin{figure}
    \centering
    \includegraphics[width=0.9\textwidth]{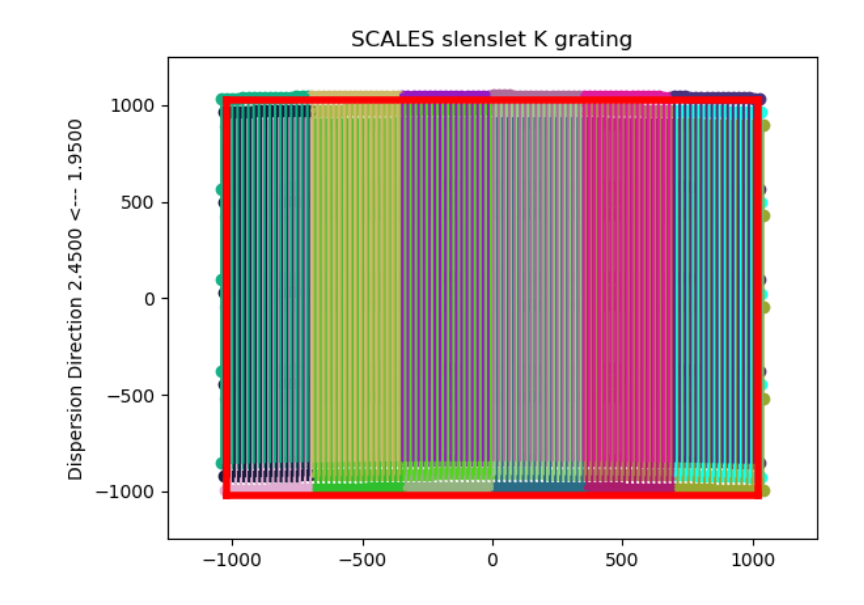}
    \caption{Spectral layout for the K-band mid-resolution mode. Different colors are used to represent spectra from different columns of spaxels.}
    \label{fig:slenslitK}
\end{figure}

\begin{figure}
    \centering
    \includegraphics[width=0.9\textwidth]{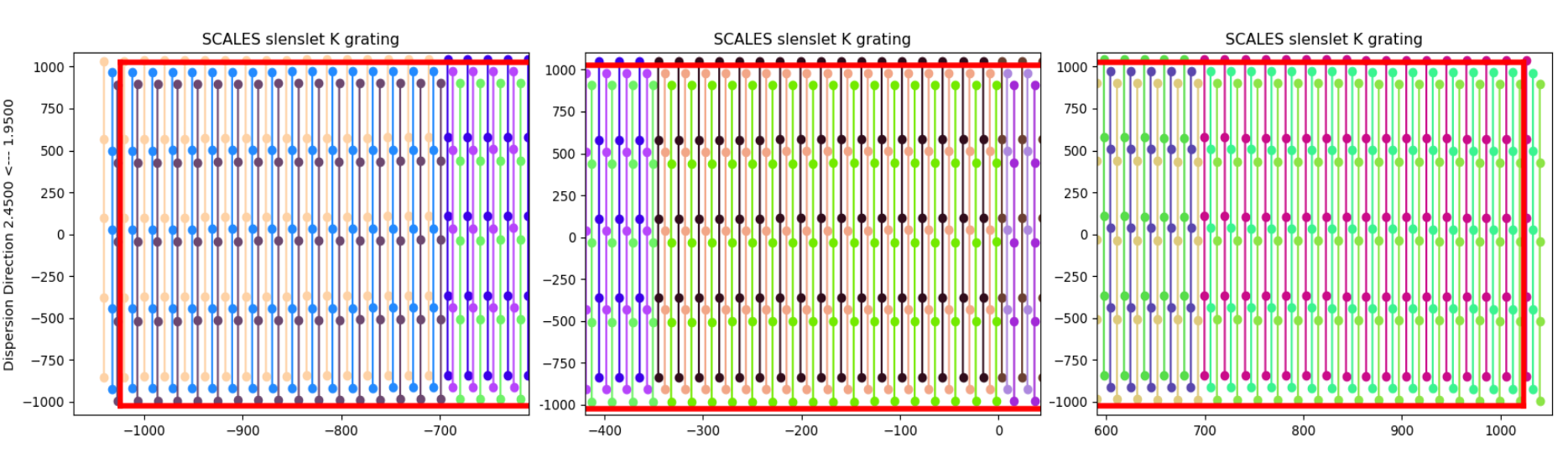}
    \caption{Zoomed spectral layout of the slenslit mode in K-band. Each frame spans approximately one-third of the detector. Maps like these provided verification of distance between spectra.}
    \label{fig:slenslitKzoom}
\end{figure}

The slenslit medium resolution mode spectral layout for the K-band passband is shown in Figures~\ref{fig:slenslitK} and \ref{fig:slenslitKzoom}. We used these maps to optimize the slenslit optical design, ensuring that each spaxel's spectra was at least six pixels from it's nearest neighbor to avoid crosstalk. These maps also allowed identification of missing data, either partial spectral passbands or entire spaxels, which varied with disperser architecture. Finally, the spectral positions coupled with the Airy radius for each wavelength directly informed the instrument simulator and development of the data reduction pipeline. 

\subsection{Predicted throughput}
\begin{figure}
    \centering
    \includegraphics[width=0.8\textwidth]{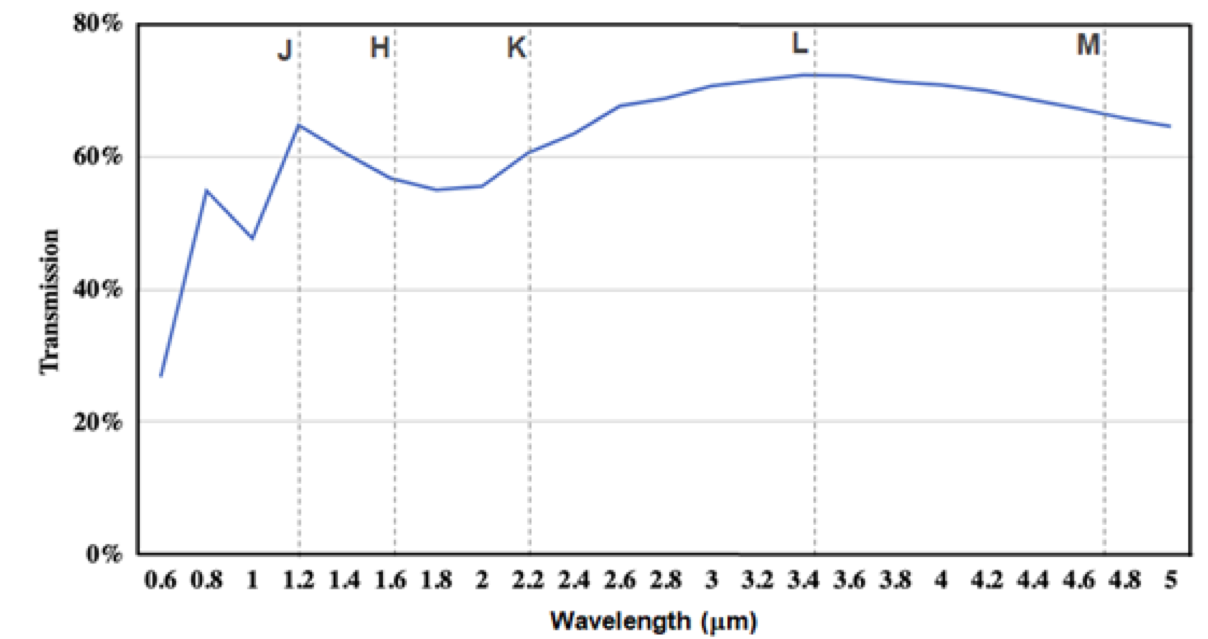}
    \caption{Imager throughput estimate}.
    \label{fig:ImagerTP}
\end{figure}

\begin{figure}
    \centering
    \includegraphics[width=0.8\textwidth]{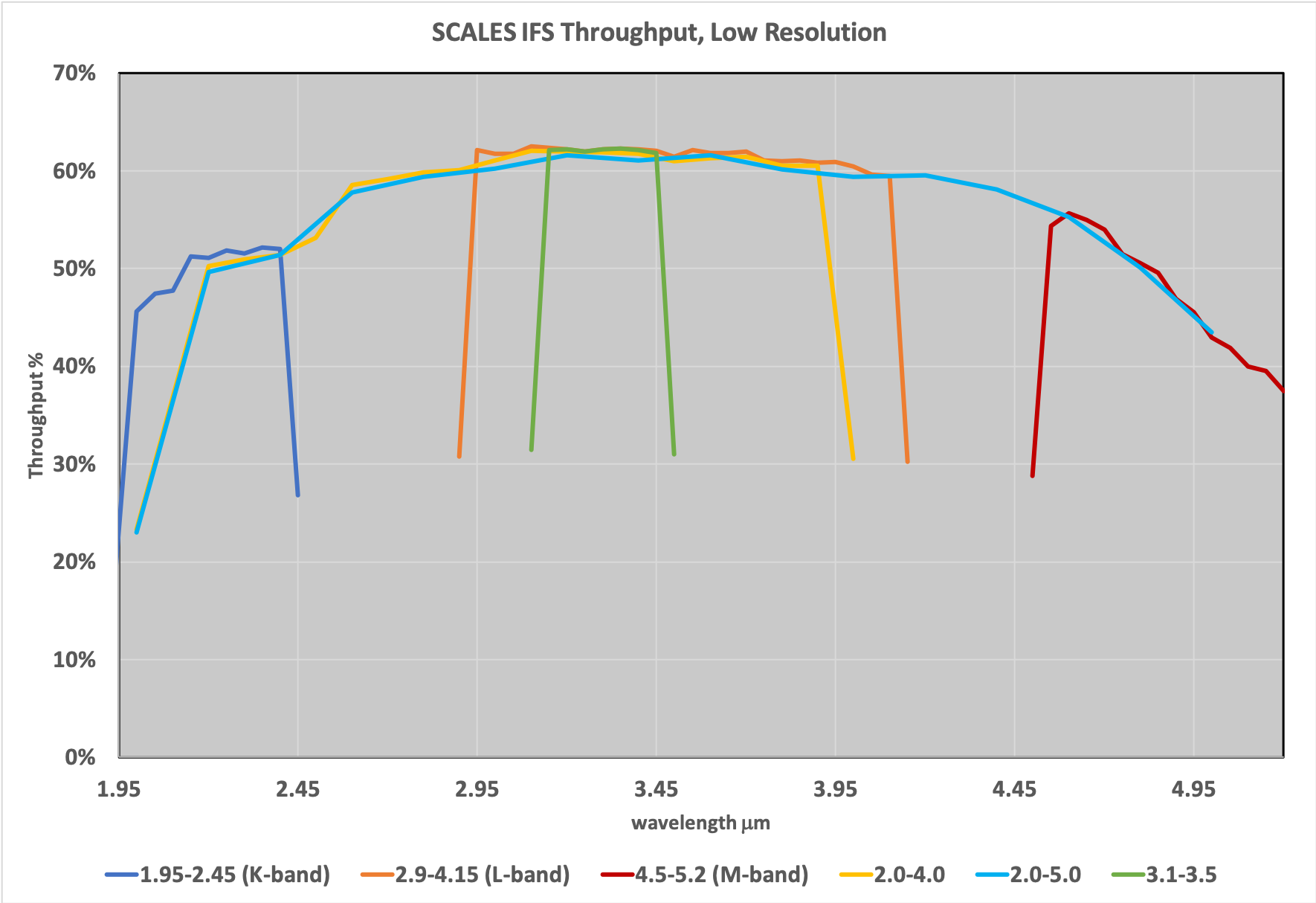}
    \caption{Low resolution spectrograph throughput estimate}.
    \label{fig:LoResTP}
\end{figure}

\begin{figure}
    \centering
    \includegraphics[width=0.8\textwidth]{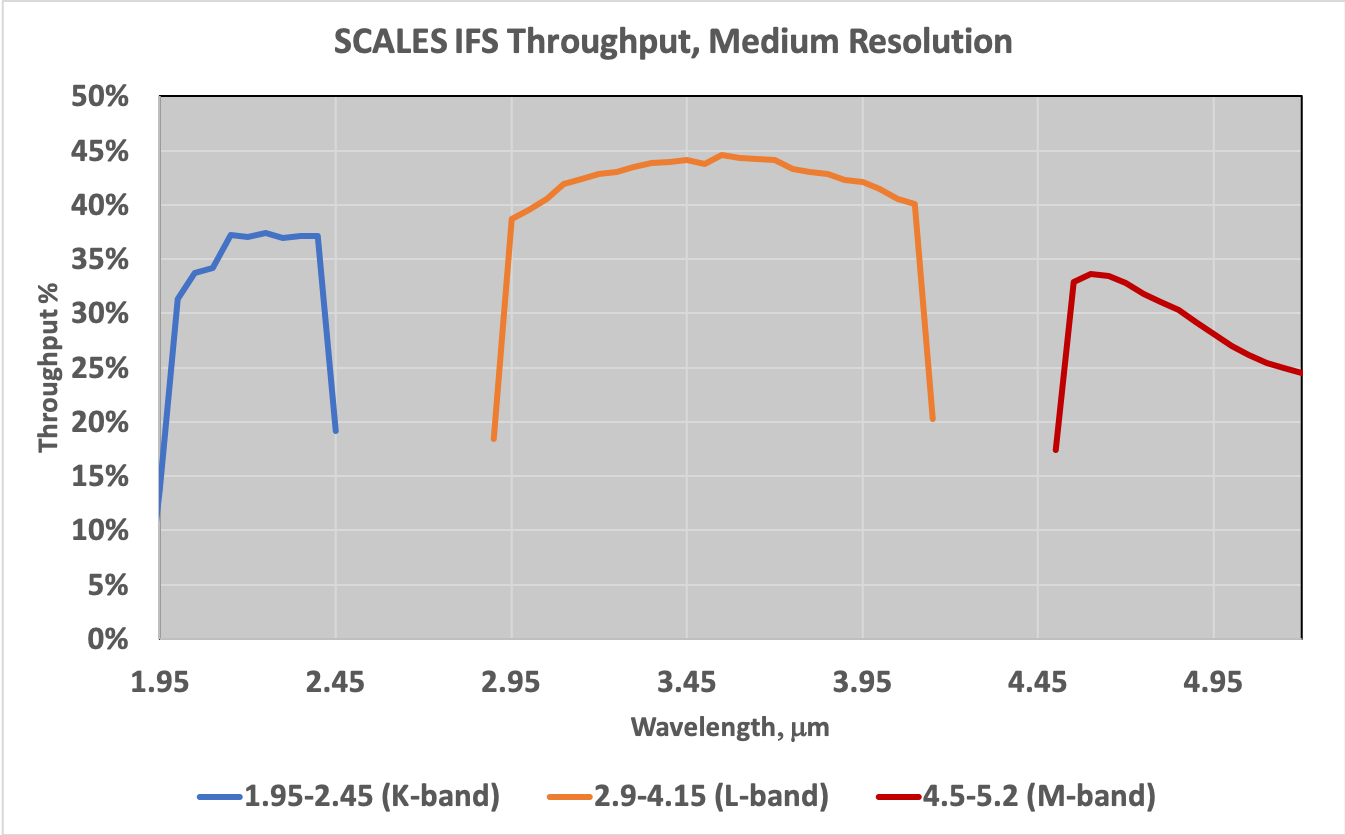}
    \caption{Medium resolution spectrograph throughput estimate}.
    \label{fig:MedResTP}
\end{figure}

A full throughput budget is under development for SCALES, for the imager and low- and medium-resolution spectrograph modes. It includes the following: Reflectivity of all gold-coated mirrors; vendor supplied filter transmission curves; vendor modeled AR coatings of refractive elements (entrance window, lenslet array and prisms); grating efficiency estimates derived from rigorous coupled wave analysis; losses from the Lyot stop and pinhole grid at the micropupil plane and measured H2RG detector quantum efficiency. It does not include atmospheric, telescope or adaptive optics losses. It also does not include scattering off of surfaces or the microlens array or degradation of coatings with time. Figures~\ref{fig:ImagerTP}, \ref{fig:LoResTP} and \ref{fig:MedResTP} show predicted throughput versus passband for the imager, low- and medium-resolution spectrograph, respectively.

\section{FABRICATION}
\begin{figure}
    \centering
    \includegraphics[width=0.8\textwidth]{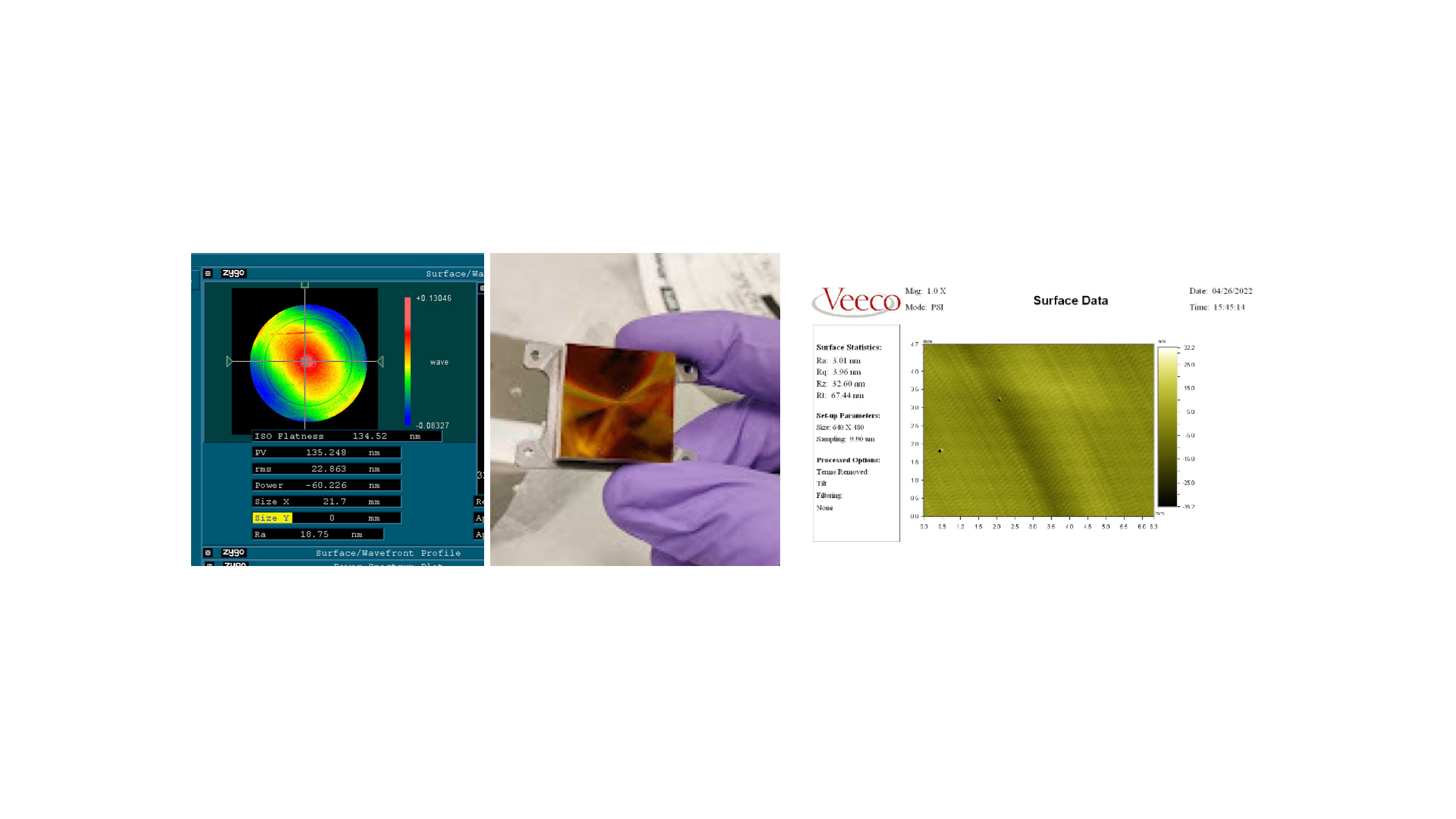}
    \caption{Flat mirror FM3 to be mounted on the piezoelectric tip/tilt stage. This mirror is Al 443 with nickel plating. Testing has begun to characterize figure error (left) and surface roughness (right)}.
    \label{fig:FM3Testing}
\end{figure}
One of the difficult engineering tasks for cryogenic instrumentation is in how one deals with large differences in thermal expansion coefficients (CTEs) from room temperature to cryogenic temperatures. This impacts both mount design and especially alignment. We chose, as much as possible, to use a single material: aluminum 6061-T6. The optical bench, almost all mechanisms (with exception of the cryogenic field steering mirror), mirror mounts and mirrors themselves (with exception of the tip/tilt mirror) will be machined out of 6061. Recent advances in single point diamond turning and polishing of special alloys of aluminum (RSA) have resulted in achievable surface roughnesses of reliably $<$ 5 nm RMS. We have performed physical optics modeling to confirm that these surface roughnesses and the expected power spectral density curve for figure error will be adequate for coronagraphy at mid-infrared wavelengths. Utilizing aluminum for the optical bench, mounts and optics allows us to align the complicated optical system at room temperature.

We have in hand the flat mirror FM3 (Figure~\ref{fig:Foreoptics}). Because it will be mounted to a piezoelectric tip/tilt stage it has been constructed out of aluminum 443 rather than 6061 (to match the materials of the stage to which it is mounted). The fine grain structure of Al 443 is not as optimal as RSA 6061 for achieving exceptional surface smoothness, so it has been nickel-plated before the final reflective gold coating. We are currently testing this mirror for reflectivity, surface roughness and surface figure both at room temperature and cryogenic temperatures to verify that it will meet SCALES tight manufacturing tolerances of $<$ 24 nm RMS wavefront error in figure and $<$ 3 nm RMS surface roughness. Figure~\ref{fig:FM3Testing} shows preliminary results from this testing, which we anticipate reporting in a future publication. 

\section{CONCLUSION}
We have reported on the progress in the optical design of the SCALES instrument, a coronagraphic mid-infrared imager and integral field spectrograph for the Keck adaptive optics system. We presented the design, baseline performance and modeling, and will in the future report on progress in fabrication, verification and testing as we work towards final commissioning of the exciting SCALES instrument. 

\section*{ACKNOWLEDGMENTS}
We are grateful to the Heising-Simons Foundation, the Alfred P. Sloan Foundation, and the Mt. Cuba Astronomical Foundation for their generous support of our efforts.  This project also benefited from work conducted under NSF Grant 1608834 and the NSF Graduate Research Fellowship Program.  In addition, we thank the Robinson family and other private supporters, without whom this work would not be possible.

\bibliography{report}
\bibliographystyle{spiebib}
\end{document}